\documentclass[aps,prd,showpacs,twocolumn,floatfix,nofootinbib]{revtex4-1}
\usepackage[dvipdfmx]{graphicx}
\usepackage{amsmath,amssymb,siunitx}
\usepackage{color,ulem}
\usepackage{graphicx}
\usepackage{bm,latexsym,amsmath,amssymb,amsfonts,mathrsfs}
\usepackage{dcolumn}   % needed for some tables
\usepackage{color}

\newcommand{\simgt}{\lower.5ex\hbox{$\; \buildrel > \over \sim \;$}}
\newcommand{\simlt}{\lower.5ex\hbox{$\; \buildrel < \over \sim \;$}}
\newcommand{\solM}{M_{\odot}}

\newcommand{\Kyoto}{\texttt{TF2+\_KyotoTidal}}
\newcommand{\NRTidal}{\texttt{TF2+\_NRTidal}}
\newcommand{\NRTidalvt}{\texttt{TF2+\_NRTidalv2}}
\newcommand{\PNTFt}{\texttt{TF2\_PNTidal}}
\newcommand{\PNTFtp}{\texttt{TF2+\_PNTidal}}
\newcommand{\TaylorFt}{\texttt{TaylorF2}}
\newcommand{\TFt}{\texttt{TF2}}
\newcommand{\TFtp}{\texttt{TF2+}}

\begin{document}
\title{Reanalysis of the binary neutron star mergers GW170817 and GW190425 using numerical-relativity calibrated waveform models}
\author{Tatsuya Narikawa$^{1,2}$}
\email{narikawa@icrr.u-tokyo.ac.jp}
\author{Nami Uchikata$^{3,2}$}
\email{uchikata@icrr.u-tokyo.ac.jp}
\author{Kyohei Kawaguchi$^{2,4}$}
\email{kkawa@icrr.u-tokyo.ac.jp}
\author{Kenta Kiuchi$^{4,5}$}
\email{kenta.kiuchi@aei.mpg.de}
\author{Koutarou Kyutoku$^{1,6,7,5}$}
\email{kyutoku@tap.scphys.kyoto-u.ac.jp}
\author{Masaru Shibata$^{4,5}$}
\email{mshibata@yukawa.kyoto-u.ac.jp}
\author{Hideyuki Tagoshi$^{2}$}
\email{tagoshi@icrr.u-tokyo.ac.jp}
\affiliation{
$^1$Department of Physics, Kyoto University, Kyoto 606-8502, Japan\\
$^2$Institute for Cosmic Ray Research, The University of Tokyo, Chiba 277-8582, Japan\\
$^3$Graduate School of Science and Technology, Niigata University, Niigata 950-2181, Japan\\
$^4$Max Planck Institute for Gravitational Physics (Albert Einstein Institute),
Am M$\ddot{u}$hlenberg 1, Potsdam-Golm 14476, Germany\\
$^5$Center for Gravitational Physics, Yukawa Institute for Theoretical
Physics, Kyoto University, Kyoto 606-8502, Japan\\
$^6$Theory Center, Institute of Particle and Nuclear Studies, KEK,
Tsukuba 305-0801, Japan\\
$^7$Interdisciplinary Theoretical and Mathematical Science (iTHEMS) Research Group, RIKEN,
Wako, Saitama 351-0198, Japan\\
}
\date{\today}

\begin{abstract}
We reanalyze gravitational waves from binary-neutron-star mergers GW170817 and GW190425 using a numerical-relativity (NR) calibrated waveform model, the \Kyoto~model, which includes nonlinear tidal terms.
For GW170817, by imposing a uniform prior on the binary tidal deformability $\tilde{\Lambda}$,
the symmetric $90\%$ credible interval of $\tilde{\Lambda}$
is estimated to be 
$481^{+436}_{-359}$ and $402^{+465}_{-279}$
for the case of $f_\mathrm{max}=1000$ and $2048~\mathrm{Hz}$, respectively, 
where $f_\mathrm{max}$ is the maximum frequency in the analysis.
We also reanalyze the event with other waveform models: 
two post-Newtonian waveform models (\PNTFt~and \PNTFtp),  
the \NRTidal~model that is
another NR calibrated waveform model,
and its upgrade, the \NRTidalvt~model.
While estimates of parameters other than $\tilde{\Lambda}$ are broadly consistent among various waveform models,
our results indicate that estimates of $\tilde{\Lambda}$ depend on waveform models.
However, the difference is smaller than the statistical error.
For GW190425, we can only obtain little information on the binary tidal deformability.
The systematic difference among the NR calibrated waveform models 
will become significant to measure $\tilde{\Lambda}$ 
as the number of detectors and events increase and sensitivities of detectors are improved.
\end{abstract}

%\pacs{04.80.Cc, 04.80.Nn, 04.50.Kd, 04.30.-w, 04.25.Nx}
%\preprint{}

\maketitle

\section{Introduction}
\label{sec:Introduction}
%%%%%%%%%%%%%%%%%%%%%%%%%%%%%%%%%%%%%%%%%%%%%%%
Binary-neutron-star (BNS) mergers are valuable laboratories for nuclear
astrophysics. Matter effects influence the orbital evolution and
gravitational radiation through the tidal interaction between the
neutron stars (NSs) in the late inspiral phase. Additionally, the presence of
material gives rise to electromagnetic emission primarily after the gravitational radiation. Because these signals depend on the
properties of nuclear matter, their observations allow us to study
nuclear-matter properties such as the equation of state (EOS) for
NS matter.

GW170817 \cite{TheLIGOScientific:2017qsa} and associated electromagnetic
counterparts are used to derive various constraints on NS
properties and the underlying EOS. The existence of a blue
component in the kilonova/macronova AT 2017gfo \cite{GBM:2017lvd}
might suggest that the merger remnant did not collapse promptly to a black
hole. Thus, the maximum mass of the NS should not be as small
as $\sim 2M_\odot$ \cite{Shibata:2017xdx} and also the radii of
high-mass NS may not be very small, e.g., the radius of the
maximum-mass configuration is likely to be larger than \SI{9.60}{\km}
\cite{Bauswein:2017vtn} (but see also Ref.~\cite{Kiuchi:2019lls}). 
At the same time, the short gamma-ray burst GRB 170817A
\cite{Monitor:2017mdv} and the absence of magnetar-powered emission in
AT 2017gfo suggest that the remnant NS collapsed early in the postmerger phase
(but see also
Refs.~\cite{Metzger:2018uni,Ai:2018jtv,Li:2018hzy,Yu:2017syg}). Accordingly,
a maximum mass of $\gtrsim 2.3M_\odot$ is also unlikely
\cite{Shibata:2017xdx,Margalit:2017dij,Rezzolla:2017aly,Ruiz:2017due,Shibata:2019ctb}.

Tidal deformability extracted via cross-correlating gravitational-wave (GW)
data of GW170817 with theoretical waveforms gives us more concrete
information about the NS than electromagnetic
counterparts. The LIGO-Virgo collaborations (LVC) initially put an upper limit on the
binary tidal deformability $\tilde{\Lambda}$ of the binary as $\tilde{\Lambda} \lesssim
800$ with the prior on the dimensionless NS spin being chosen to
be $|\chi| \leq 0.05$ \cite{TheLIGOScientific:2017qsa}. This limit
is later corrected to be
$\tilde{\Lambda} \lesssim 900$ in Ref.~\cite{Abbott:2018wiz}, where the
result of updated analysis is also reported as, e.g., $\tilde{\Lambda} =
300^{+420}_{-230}$ for a particular set of assumptions. The constraint
can be further improved by assuming the EOS to be common
for both NS \cite{Abbott:2018exr, LIGOScientific:2019eut} (but see also Ref.~\cite{Kastaun:2019bxo}) as is also done in an
independent analysis \cite{De:2018uhw, Capano:2019eae}. These constraints are used to
investigate the NS EOS
\cite{Fattoyev:2017jql,Annala:2017llu,Raithel:2018ncd} as well as those
for quark and hybrid stars
\cite{Zhou:2017pha,Paschalidis:2017qmb,Nandi:2017rhy}. While it has been
claimed based on a limited number of numerical-relativity (NR) simulations
that $\tilde{\Lambda} \gtrsim 400$ is necessary to account for the
ejecta mass of $\approx 0.05M_\odot$ required to explain AT 2017gfo
\cite{Radice:2017lry} 
(see Ref. \cite{Radice:2018ozg} for an updated analyses stating the bound on the tidal deformability in $323\leq \tilde{\Lambda} \leq 776$ and Ref. \cite{Coughlin:2018fis}, stating the similar bound to be in $302\leq \tilde{\Lambda} \leq 860$), 
later systematic investigations reveal that this
argument is premature \cite{Kiuchi:2019lls}.

Recently, the discovery of the second BNS merger event, GW190425, was reported~\cite{Abbott:2020uma}.
This binary system is massive and it is intrinsically difficult to measure the tidal effect.
LVC have reported that GW190425 constrains $\tilde{\Lambda}$ to be below 650 
for the low-spin prior ($|\chi| \leq 0.05$), 
after reweighting the posterior to derive approximately the result corresponding to a flat prior in $\tilde{\Lambda}$~\cite{Abbott:2020uma}.
While GW190425 does not carry novel information on the NS properties,
multi-messenger constraints on the NS EOS have been studied \cite{Dietrich:2020lps,Landry:2020vaw}.

An accurate theoretical waveform template is crucial to extract accurately the tidal deformability of NSs from the observed GW data. For the early stage of the inspiral, the waveforms including the linear-order tidal effects derived by post-Newtonian (PN) calculation are useful~\cite{Flanagan:2007ix,Vines:2011ud}. However, the PN expansion becomes invalid as the orbit becomes relativistic, and thus, the error of the waveform becomes large in the late stage~\cite{Favata:2013rwa,Yagi:2013baa,Wade:2014vqa,Lackey:2014fwa}. Such errors would cause the systematic bias in the parameter estimation, and it would be in particular problematic for estimating the tidal deformability because the tidal effects on the waveform become most significant just before the merger~\cite{Hinderer:2009ca,Damour:2012yf}. The effective-one-body (EOB) formalism can solve this problem by incorporating the higher-order PN correction by re-summation techniques and calibrating them to NR waveforms~\cite{Damour:2009wj,Damour:2012yf,Bini:2012gu,Bini:2014zxa,Bernuzzi:2014owa,Hinderer:2016eia}. 
Hence, employing waveform models with the higher-order PN correction calibrated to NR waveforms, 
such as the EOB formalism, is crucial for the data analysis.

Another approach to solve the problem is to adopt phenomenological models calibrated also to NR waveforms.
Dietrich {\it et al.}~have derived a gravitational waveform model, \texttt{NRTidal}, for BNSs based on high-precision NR simulations~\cite{Dietrich:2017aum}. Improved reanalyses of GW170817 with more sophisticated waveform models calibrated to NR simulation of BNS merger have been performed employing such a model~\cite{Abbott:2018wiz}. Indeed, it is pointed out that the value of the tidal deformability tends to be overestimated if the PN models are employed for the parameter estimation~\cite{TheLIGOScientific:2017qsa}. 
Recently, its upgrade, the \texttt{NRTidalv2} model, which is calibrated to more precise NR waveforms, has been derived~\cite{Dietrich:2019kaq}.
Kawaguchi {\it et al.}~have also developed a model (hereafter the \texttt{KyotoTidal}~model) for frequency-domain gravitational waveforms of inspiraling BNSs~\cite{Kawaguchi:2018gvj}. 
In particular, this model is derived independently from the \texttt{NRTidal} model employing different NR waveforms. 
Since the \texttt{NRTidal} model is to date the only NR calibrated waveform model that is used for parameter estimation of GWs from BNS mergers, the analysis comparing these three NR calibrated waveform models would help us to understand the systematic differences in resulting constraints on tidal deformability.

In this paper, we reanalyze the data around GW170817 and GW190425 against a NR calibrated waveform model, the \Kyoto~model and 
present constraints on the binary tidal deformability.
We also reanalyze the events with other waveform models: 
two PN (\PNTFt~and \PNTFtp), \NRTidal, and \NRTidalvt~models.
Here, \texttt{TF2} is the abbreviation of \texttt{TaylorF2}, which is the PN waveform model 
for a point-particle part~\cite{Buonanno:2009zt,Blanchet:2013haa}
and \texttt{TF2+}~\cite{Kawaguchi:2018gvj} is a phenomenologically extended model of \texttt{TF2}.

The remainder of this paper is organized as follows.
In Sec.~\ref{sec:PE}, we explain the methods for parameter estimation including waveform models used to reanalyze GW170817 and GW190425.
In Sec.~\ref{sec:results}, we present results of our analysis of GW170817 and a comparison of our analysis with the  LVC analysis.
In sec. \ref{sec:GW190425_EOS}, we present results of GW190425 and discuss constraints on NS EOS by combining information obtained from GW170817 and GW190425.
Section~\ref{sec:summary} is devoted to a summary. 
In Appendix, we present an in-depth study of our results for GW170817 by separate analysis for the LIGO twin detectors to interpret the origin of the complex structure at the high-$\tilde{\Lambda}$ region for the posterior probability density function (PDF) of $\tilde{\Lambda}$ (see also Ref.~\cite{Narikawa:2018yzt}).
Unless otherwise stated, we employ the units $c=G=1$, where $c$ and $G$ are the speed of light and the gravitational constant, respectively.

%%%%%%%%%%%%%%%%%%%%%%%%%%%%%%%%%%%%%%%%%%%%%%%
\section{Parameter estimation methods}
\label{sec:PE}
%%%%%%%%%%%%%%%%%%%%%%%%%%%%%%%%%%%%%%%%%%%%%%%
%%%%%%%%%%%%%%%%%%%%%%%%%%%%%%%%%%%%%%%%%%%%%%%
\subsection{Data and Bayesian inference}
\label{sec:methods}
%%%%%%%%%%%%%%%%%%%%%%%%%%%%%%%%%%%%%%%%%%%%%%%
We use Bayesian inference to reanalyze GW170817 and GW190425 with 
various waveform models that incorporate tidal effects in a different manner.
Our analysis follows the one performed in our recent work \cite{Narikawa:2018yzt}, and uses the public data by LVC\footnote{\url{https://www.gw-openscience.org/catalog/GWTC-1-confident/single/GW170817/}
for Hanford and Virgo for GW170817, \url{https://dcc.ligo.org/LIGO-T1700406/public}
for Livingston for GW170817, and \url{https://dcc.ligo.org/LIGO-T1900685/public} for GW190425}.
We calculate the posterior PDF, $p(\vec{\theta}|\vec{s}(t),H)$, 
for the binary parameters $\vec{\theta}$ for the gravitational waveform model, $H$, 
given the LIGO Hanford, LIGO Livingston, and Virgo data $\vec{s(t)}$ via
\begin{eqnarray}
 p(\vec{\theta}|\vec{s}(t),H)\propto p(\vec{\theta}|H)p(\vec{s}(t)|\vec{\theta},H).
\end{eqnarray}
$p(\vec{\theta}|H)$ is the prior for the binary parameters.
The likelihood $p(\vec{s}(t)|\vec{\theta},H)$ is evaluated by assuming stationarity and Gaussianity for the detector noise using the noise power spectrum density derived with BayesLine\footnote{\url{https://dcc.ligo.org/LIGO-P1900011/public}}.
We compute PDFs
by using stochastic sampling engine based on nested sampling~\cite{Skilling:2006,Veitch:2009hd}.
Specifically, we use the parameter estimation software, LALInference~\cite{Veitch:2014wba, lalinference},
which is one of the software of LIGO Algorithm Library (LAL) software suite. 
We take the frequency range from 23 Hz for GW170817 and 19.4 Hz for GW190425 to $f_\mathrm{max}$.
Here, the maximum frequency $f_\mathrm{max}$ is chosen from two values, 1000 Hz or min[$f_\mathrm{ISCO}$,~$f_s/2$], where $f_\mathrm{ISCO}$ is twice the orbital frequency at the innermost stable circular orbit of a Schwarzschild black hole with total mass of the binary, and $f_s$ is the sampling rate of data. We set $f_s=4096$ Hz.
The former choice is made because the \Kyoto~model is calibrated in the frequency range of 10--1000 Hz.
The latter choice corresponds to the assumption that the inspiral stage is terminated at the smaller of $f_\mathrm{ISCO}$ and $f_s/2$.
In this work, we represent the latter choice by $f_\mathrm{max} = 2048~\mathrm{Hz}$ for simplicity.

\begin{widetext}

\begin{table}[htbp]
%\begin{center}
%\begin{tabular}{cc}
%\begin{minipage}{0.3\hsize}
\begin{center}
\begin{tabular}{c|cc|cc}
\hline \hline
Model name & Point-particle part &           & Tidal part          &            \\ \hline
           & Amplitude      & Phase & Amplitude & Phase \\ \hline
\PNTFt & 3PN & 3.5PN & 5+1PN & 5+2.5PN \\ 
\PNTFtp & 6PN & 6PN & 5+1PN & 5+2.5PN \\ 
\Kyoto & 6PN & 6PN & Polynomial & Nonlinear \\ 
\NRTidal & 6PN & 6PN & - & $\mathrm{Pad\acute{e}}$ approximation \\ 
\NRTidalvt & 6PN & 6PN & $\mathrm{Pad\acute{e}}$ approximation & $\mathrm{Pad\acute{e}}$ approximation \\
\hline \hline
\end{tabular}
\caption{
Waveform models used to reanalyze GW170817 and GW190425.
Our reference model, the \Kyoto~model incorporates \TFtp, as the point-particle and spin parts, and 
NR calibrated tidal effects.
The \texttt{TF2} approximant employs the 3.5PN- and 3PN-order formulas for the phase and amplitude, respectively as the point-particle part,
and
treats aligned spins and
incorporates 3.5PN-order formula in spin-orbit interactions,
2PN-order formula in spin-spin, 
and self-spin interactions.
\TFtp~is the \texttt{TF2} approximant supplemented with phenomenological higher-order PN terms calibrated to \texttt{SEOBNRv2} for the point-particle part.
The \NRTidal~model is another model whose tidal effects are calibrated to NR.
The \NRTidalvt~model is the upgrade of the \NRTidal~model.
The \PNTFt~and \PNTFtp~models employ the PN tidal-part phase formula.
}
\label{table:TidalWFs}
\end{center}
%\end{minipage}
\end{table}

\end{widetext}

%----------------------------------------------------------------%
\begin{figure}[htbp]
  \begin{center}
    \includegraphics[keepaspectratio=true,height=60mm]{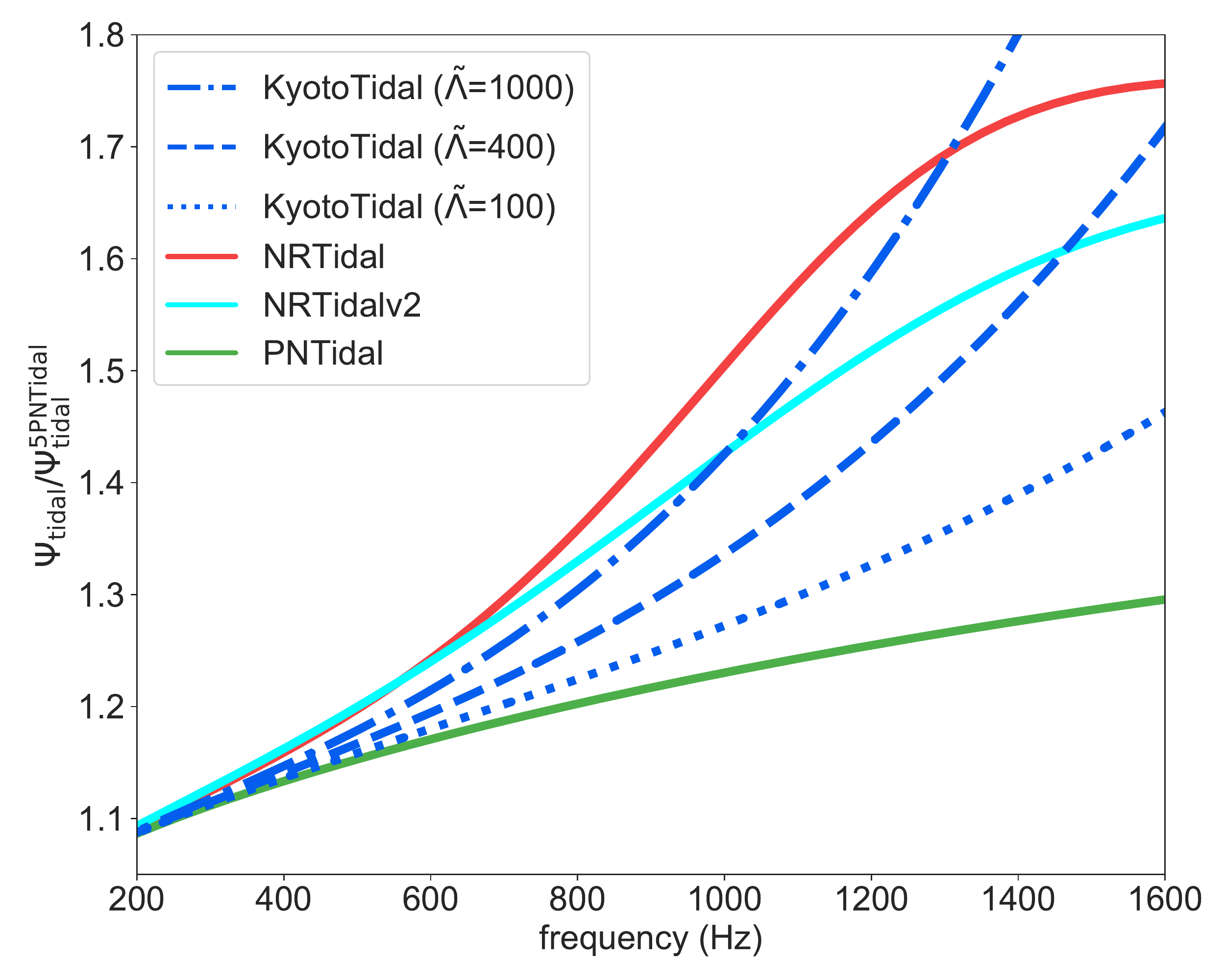}\\
  \caption{
Tidal phase in the frequency domain normalized by the leading, Newtonian (relative 5PN-order) tidal phase formula.
Here, we use $(m_1,~m_2)=(1.35M_{\odot},~1.35M_{\odot})$. 
We show $\tilde{\Lambda} = 1000$ (dot-dashed, blue), 400 (dashed, blue), and 100 (dotted, blue) for the \texttt{KyotoTidal} model.
The \texttt{NRTidal} model (solid, red), the \texttt{NRTidalv2} model (solid, cyan), 
and the 5+2.5PN-order tidal-part phase formula, \texttt{PNTidal} (solid, green),
are also presented, 
which are independent of $\tilde{\Lambda}$ when normalized by the leading tidal phase.
}%
\label{fig:TidalWFs}
\end{center}
\end{figure}
%----------------------------------------------------------------%

%%%%%%%%%%%%%%%%%%%%%%%%%%%%%%%%%%%%%%%%%%%%%%%
\subsection{Waveform models for inspiraling BNSs}
\label{sec:WF}
%%%%%%%%%%%%%%%%%%%%%%%%%%%%%%%%%%%%%%%%%%%%%%%

We use various analytic frequency-domain waveform models for the inspiral phase, all of which are constructed based on the PN formulas.
The features of each waveform model are summarized in Table \ref{table:TidalWFs}.
The Fourier transform of the gravitational waveform can be written as
\begin{eqnarray}
 \tilde{h}(f) = A(f) e^{i\Psi(f)},
\end{eqnarray}
where the amplitude $A(f)$ and the phase $\Psi(f)$ can be decomposed into the point-particle evolution, the spin effects, and the tidal effects as
\begin{eqnarray}
 A(f) = A_\mathrm{point-particle}(f) + A_\mathrm{spin}(f) + A_\mathrm{tidal}(f),
\end{eqnarray}
and
\begin{eqnarray}
 \Psi(f) = \Psi_\mathrm{point-particle}(f) + \Psi_\mathrm{spin}(f) + \Psi_\mathrm{tidal}(f).
\end{eqnarray}

We use \texttt{TF2}~\cite{Buonanno:2009zt,Blanchet:2013haa} and phenomenologically extended model of \TFt, called \TFtp~(see Ref.~\cite{Kawaguchi:2018gvj} and below) as BBH baseline, which consists of point-particle and spin parts.
Here, the 3.5PN-order formula for the phase and 3PN-order formula for the amplitude are employed as the point-paticle part of \TFt~\cite{Khan:2015jqa}.
For \TFtp, both the phase and amplitude of the point-particle part are extended to the 6PN-order
by fitting the \texttt{SEOBNRv2} model~\cite{Purrer:2015tud,Taracchini:2013rva}.

All waveform models used in our parameter estimation analyses
assume that the spins of component stars are aligned with the orbital angular momentum,
and incorporate 3.5PN-order formula in couplings between the orbital angular momentum and the component spins~\cite{Bohe:2013cla},
2PN-order formula in point-mass spin-spin, 
and self-spin interactions~\cite{Arun:2008kb, Mikoczi:2005dn}.

During the BNS inspiral, at the leading order,
the induced quadrupole moment tensor $Q_{ij}$ is proportional to 
the external tidal field tensor ${\cal E}_{ij}$ as $Q_{ij} = -\lambda {\cal E}_{ij}$.
The information about the NS EOS can be quantified by the tidal deformability parameter $\lambda$~\cite{Flanagan:2007ix, Hinderer:2007mb}.
The leading order tidal contribution to the GW phase evolution (relative 5PN-order) is governed by the symmetric contribution of NS tidal deformation, characterized by the binary tidal deformability~\cite{Flanagan:2007ix}
\begin{eqnarray}
 \tilde{\Lambda} = \frac{16}{13} \frac{(m_1+12m_2)m_1^4\Lambda_1+(m_2+12m_1)m_2^4\Lambda_2}{(m_1+m_2)^5},
\end{eqnarray}
which is a mass-weighted linear combination of the tidal deformability of the both components,
where $m_{1,2}$ is the component mass and 
$\Lambda_{1,2}$ is the dimensionless tidal deformability parameter of each star 
$\Lambda_{1,2} = \lambda_{1,2} / m_{1,2}^5$.
The antisymmetric contribution $\delta \tilde{\Lambda}$ terms introduced in Refs.~\cite{Favata:2013rwa,Wade:2014vqa} are always subdominant on the tidal effects to the GW phase and 
the symmetric contribution $\tilde{\Lambda}$ terms dominate~\cite{Favata:2013rwa,Wade:2014vqa}.
In this paper, we ignore the $\delta\tilde{\Lambda}$ contribution.

The \PNTFt~ and \PNTFtp~models denote the waveform models employing \TFt~and \TFtp~as the BBH baseline, respectively.
Both the \PNTFt~and the \PNTFtp~models employ the 2.5PN-order (relative 5+2.5PN-order) tidal-part phase formula~\cite{Damour:2012yf}
\begin{eqnarray}
 &&\Psi_\mathrm{tidal}^\mathrm{PNTidal} = \frac{3}{128\eta} \left[ -\frac{39}{2} \tilde{\Lambda}  \right. x^{5/2} \nonumber \\
&&\times \left. \left( 1 + \frac{3115}{1248} x - \pi x^{3/2} + \frac{28024205}{3302208} x^2 - \frac{4283}{1092} \pi x^{5/2} \right) \right], \nonumber \\
\end{eqnarray}
where $x=[\pi M_\mathrm{tot} (1+z) f]^{2/3}$ is the dimensionless PN parameter, $M_\mathrm{tot} = m_1+m_2$ is the total mass, $\eta=m_1 m_2 / (m_1+m_2)^2$ is the symmetric mass ratio, and $z$ is the source redshift.
The tidal-part amplitude for both \PNTFt~and \PNTFtp~models employ the 5+1PN-order amplitude formula given by~\cite{Damour:2012yf}
\begin{eqnarray}
 A_\mathrm{tidal}^\mathrm{PNTidal} &=& \sqrt{\frac{5 \pi \eta}{24}} \frac{M_\mathrm{tot}^2 (1+z)^2}{d_L} \tilde{\Lambda} x^{-7/4} \nonumber \\
 &\times& \left( - \frac{27}{16} x^5 - \frac{449}{64} x^6 \right),
\label{eq:PNTidal_amp}
\end{eqnarray}
where $d_L$ is the luminosity distance to the source.

The \Kyoto~model is a NR calibrated waveform model
for the inspiral phase of BNS mergers~\cite{Kiuchi:2017pte,Kawaguchi:2018gvj}.
The \Kyoto~model employs \TFtp~as the BBH baseline and 
extends the 2.5PN-order (relative 5+2.5PN-order) tidal-part phase formula~\cite{Damour:2012yf}
by multiplying $\tilde{\Lambda}$ by a nonlinear correction
to model the tidal part of the GW phase. 
The functional forms of the tidal-part phase is 
\begin{eqnarray}
 &&\Psi_\mathrm{tidal}^\mathrm{KyotoTidal} = \frac{3}{128\eta} \left[ -\frac{39}{2} \tilde{\Lambda} \left( 1+a \tilde{\Lambda}^{2/3} x^p \right) \right] x^{5/2} \nonumber \\
&&\times \left( 1 + \frac{3115}{1248} x - \pi x^{3/2} + \frac{28024205}{3302208} x^2 - \frac{4283}{1092} \pi x^{5/2} \right), \nonumber \\
\end{eqnarray}
where $a=12.55$ and $p=4.240$.
The tidal-part amplitude is extended by adding 
the higher-order PN tidal effects to Eq.~(\ref{eq:PNTidal_amp}) as
\begin{eqnarray}
 A_\mathrm{tidal}^\mathrm{KyotoTidal} &=& \sqrt{\frac{5 \pi \eta}{24}} \frac{M_\mathrm{tot}^2 (1+z)^2}{d_L} \tilde{\Lambda} x^{-7/4} \nonumber \\
 &\times& \left( - \frac{27}{16} x^5 - \frac{449}{64} x^6 - b x^{r} \right),
\end{eqnarray}
where $b=4251$ and $r=7.890$.
In the \texttt{KyotoTidal} model, the hybrid waveforms constructed from high-precision NR waveforms and the \texttt{SEOBNRv2T} waveforms~\cite{Purrer:2015tud,Taracchini:2013rva,Hinderer:2016eia,Steinhoff:2016rfi,Lackey:2018zvw} are used for model calibration in the frequency range of 10--1000 Hz. 
The phase difference between the \Kyoto~model and the hybrid waveforms is smaller than 0.1 rad up to 1000 Hz for $300\lesssim \tilde{\Lambda} \lesssim1900$ and for the mass ratio $q=m_2/m_1\leq1$ between 0.73 and 1~\cite{Kawaguchi:2018gvj}.
In Ref.~\cite{Kawaguchi:2018gvj}, it is shown that the mismatch between the \Kyoto~model and the hybrid waveforms is always smaller than $1.1 \times 10^{-5}$ in the frequency range of 10--1000 Hz.

The \texttt{NRTidal} model is another approach to describe tidal effects calibrated to NR waveforms~\cite{Dietrich:2017aum}.
The \NRTidal~model employs \TFtp~as the BBH baseline. 
For the tidal effects, this model
extends the linear-order effects by effectively adding the higher-order PN terms of the tidal contribution to the GW phase.
As shown in Ref.~\cite{Dietrich:2017aum}, the expression of the tidal phase is given by the form of a rational function:
\begin{eqnarray}
 \Psi_\mathrm{tidal}^\mathrm{NRTidal} &=& \frac{3}{128\eta} \left[ -\frac{39}{2} \tilde{\Lambda} x^{5/2} \right. \nonumber \\
&\times& \left. \frac{1 + \tilde{n}_1 x + \tilde{n}_{3/2} x^{3/2} + \tilde{n}_2 x^2 + \tilde{n}_{5/2} x^{5/2}}{1 + \tilde{d}_1 x + \tilde{d}_{3/2} x^{3/2}} \right], \nonumber \\
\end{eqnarray}
where $\tilde{d}_1 = \tilde{n}_1 - 3115/1248$, the other parameters are $(\tilde{n}_1,~\tilde{n}_{3/2},~\tilde{n}_2,~\tilde{n}_{5/2}) = (-17.428,~31.867,~-26.414,~62.362)$ and $\tilde{d}_{3/2} = 36.089$.
We do not consider the tidal-part amplitude for this model following the original form~\cite{Dietrich:2017aum}.

The \NRTidalvt~model is an upgrade of the \NRTidal~model~\cite{Dietrich:2019kaq}.
Specifically, they derive a new expression for the tidal phase which is calibrated to more accurate NR waveforms as
\begin{eqnarray}
 &&\Psi_\mathrm{tidal}^\mathrm{NRTidalv2} = \frac{3}{128\eta} \left[ -\frac{39}{2} \tilde{\Lambda} x^{5/2} \right. \nonumber \\
&&\times \left. \frac{1 + \tilde{n}'_1 x + \tilde{n}'_{3/2} x^{3/2} + \tilde{n}'_2 x^2 + \tilde{n}'_{5/2} x^{5/2} + \tilde{n}'_3 x^{3}}{1 + \tilde{d}'_1 x + \tilde{d}'_{3/2} x^{3/2} + \tilde{d}'_2 x^{2}} \right], \nonumber \\
\label{eq:NRTidalv2_phase}
\end{eqnarray}
with $\tilde{n}'_1=\tilde{c}'_1 + \tilde{d}'_1$, 
$\tilde{n}'_{3/2} = (\tilde{c}'_1 \tilde{c}'_{3/2} - \tilde{c}'_{5/2} - \tilde{c}'_{3/2} \tilde{d}'_1 + \tilde{n}'_{5/2}) / \tilde{c}'_1$, 
$\tilde{n}'_2 = \tilde{c}'_2 + \tilde{c}'_1 \tilde{d}'_1 + \tilde{d}'_2$,
$\tilde{d}'_{3/2} = - (\tilde{c}'_{5/2} + \tilde{c}'_{3/2} \tilde{d}'_1 - \tilde{n}'_{5/2}) / \tilde{c}'_1$,
where the known coefficients are 
$\tilde{c}'_1 = 3115 / 1248$,
$\tilde{c}'_{3/2} = - \pi$,
$\tilde{c}'_2 = 28024205 / 3302208$,
$\tilde{c}'_{5/2} = - 4283 \pi / 1092$,
and the fitting coefficients are
$\tilde{n}'_{5/2} = 90.550822$,
$\tilde{n}'_3 = - 60.253578$,
$\tilde{d}'_1 = - 15.111208$,
$\tilde{d}'_2 = 8.0641096$.
They also introduce the tidal amplitude,
\begin{eqnarray}
 A_\mathrm{tidal}^\mathrm{NRTidalv2} &=& \sqrt{\frac{5 \pi \eta}{24}} \frac{M_\mathrm{tot}^2 (1+z)^2}{d_L} \tilde{\Lambda} x^{-7/4} \nonumber \\
 &\times& \left( - \frac{27}{16} x^5 \right) 
 \frac{1 + \frac{449}{108} x + \frac{22672}{9} x^{2.89}}{1 + d x^{4}},
\label{eq:NRTidalv2_amp}
\end{eqnarray}
where $d = 13477.8$.
Although the new phase model, \texttt{NRTidalv2}, introduced in Ref. \cite{Dietrich:2019kaq} includes
higher order spin-squared and spin-cubed terms with their
associated spin-induced moments, we do not add them in this work.

In Fig.~\ref{fig:TidalWFs}, we show differences in the phase evolution of tidal part among the \texttt{KyotoTidal}, \texttt{NRTidal}, \texttt{NRTidalv2}, and \texttt{PNTidal}~models.
A difference in the treatment of the tidal effects makes different $\tilde{\Lambda}$-dependence.
The tidal phase normalized by the leading (relative 5PN-order) tidal phase formula
for the \texttt{KyotoTidal} model depends on the binary tidal deformability $\tilde{\Lambda}$ 
due to the nonlinear correction.
Since the \texttt{NRTidal}, \texttt{NRTidalv2}, and \texttt{PNTidal} models employ the linear-order effects of the tidal deformability,
they are independent of $\tilde{\Lambda}$ when normalized by the leading tidal effect.
Figure~\ref{fig:TidalWFs} shows good agreement between the \Kyoto~model and the \NRTidalvt~model 
for $\tilde{\Lambda}\simeq1000$ below $1000~\mathrm{Hz}$ as suggested in Ref.~\cite{Dietrich:2019kaq}.
The \texttt{NRTidal} model gives the largest phase shift, 
the second is the \texttt{NRTidalv2} model, the third is the \texttt{KyotoTidal} model, and the \texttt{PNTidal} model gives the smallest,
for $\tilde{\Lambda}\leq1000$, up to $\sim$1000 Hz.
The \Kyoto~model is calibrated only up to 1000 Hz and overestimates tidal effects
at frequencies above $1000~\mathrm{Hz}$.
The \texttt{KyotoTidal} model gives the largest phase shift 
at frequency above 1200 Hz for $\tilde{\Lambda}=1000$,
and 
larger phase shift than the one for the \texttt{NRTidalv2} model 
at frequency above about 1000 and 1400 Hz
for $\tilde{\Lambda}=1000$ and 400, respectively.

%%%%%%%%%%%%%%%%%%%%%%%%%%%%%%%%%%%%%%%%%%%%%%%
\subsection{Source parameters}
\label{sec:parameters}
%%%%%%%%%%%%%%%%%%%%%%%%%%%%%%%%%%%%%%%%%%%%%%%
The source parameters and their prior probability distributions are chosen to follow 
those adopted in our recent work \cite{Narikawa:2018yzt},
and we mention specific choices made in this work.

For GW170817, we fix the sky location to the position of AT 2017gfo, which is an electromagnetic counterpart of GW170817~\cite{Soares-Santos:2017lru}, and estimates of the remaining source parameters.
Specifically, we estimate the luminosity distance to the source $d_L$, the binary inclination $\theta_\mathrm{JN}$, which is the angle between the total angular momentum and the line of sight, 
the polarization angle $\psi$, the coalescence time $t_c$, the phase at the coalescence time $\phi_c$,
component masses $m_{1,2}$, where we assume $m_1\geq m_2$,
the orbit-aligned dimensionless spin components of the stars $\chi_{1,2}$ 
where $\chi_{1,2} = c S_{1,2}/(G m_{1,2}^2)$ is the orbit-aligned dimensionless spin components of the stars with $S_{1,2}$ are the magnitudes of the spin angular momenta of the components,
and the binary tidal deformability $\tilde{\Lambda}$.

For GW170817, we assume a uniform distribution as the detector-frame component mass prior $m_{1,2}\sim U[0.83,~7.7]M_\odot$ with an additional constraint on the detector-frame chirp mass ${\cal M}^\mathrm{det}:=\mathcal{M} (1+z)\sim U[1.184,~2.168]M_\odot$, 
where the chirp mass is the best estimated mass parameter defined by 
${\cal M} = (m_1 m_2)^{3/5}(m_1+m_2)^{-1/5}$.
The prior range for ${\cal M}^\mathrm{det}$ is the same as that used for LVC analysis~\cite{Abbott:2018wiz}.
The impact of wider prior range for ${\cal M}^\mathrm{det}$
on parameter estimation is negligible.
We assume a uniform prior on the spin magnitudes and
we enforce $\chi_{1,2}\sim U[-0.05,~0.05]$.
This prior range of spin is 
consistent with the observed population of known BNSs that will merge within the Hubble time~\cite{Burgay:2003jj,Stovall:2018ouw},
and is referred to as the low-spin prior for the LVC analysis~\cite{Abbott:2018wiz}.
We assume a uniform prior on the binary tidal deformability, with $\tilde{\Lambda}\sim U[0,~3000]$.

For GW190425, we also estimate the sky location of the source with an isotropic prior.
We assume the detector-frame component mass prior $m_{1,2}\sim U[1.0,~5.0]M_\odot$ and
the spin and the binary tidal deformability priors are the same as the ones for GW170817.

%%%%%%%%%%%%%%%%%%%%%%%%%%%%%%%%%%%%%%%%%%%%%%%
\section{Results of GW170817}
\label{sec:results}
%%%%%%%%%%%%%%%%%%%%%%%%%%%%%%%%%%%%%%%%%%%%%%%
%%%%%%%%%%%%%%%%%%%%%%%%%%%%%%%%%%%%%%%%%%%%%%%
\subsection{Source properties other than the tidal deformability}
\label{sec:otherpara}
%%%%%%%%%%%%%%%%%%%%%%%%%%%%%%%%%%%%%%%%%%%%%%%
In this subsection, we show validity of our analysis as a sanity check by comparison with the LVC results.
Figure~\ref{fig:all_Low} shows the marginalized posterior PDFs of parameters other than the tidal deformability for various waveform models for $f_\mathrm{max} = 1000~\mathrm{Hz}$.
Table \ref{table:all_Low} presents the 90\% credible intervals 
of the luminosity distance $d_L$, the binary inclination $\theta_\mathrm{JN}$, mass parameters (the component masses $m_{1,2}$, the detector-frame chirp mass ${\cal M}^\mathrm{det}$, the source-frame chirp mass ${\cal M}$, the total mass $M_\mathrm{tot}$, and the mass ratio $q$),
and the effective spin parameter $\chi_\mathrm{eff} = (m_1 \chi_1 + m_2 \chi_2)/M_\mathrm{tot}$, which is the most measurable combination of spin components,
estimated using various waveform models~\cite{Ajith:2009bn, Racine:2008qv}. 
The source-frame chirp mass is derived by assuming a value of the Hubble constant $H_0 = 69~\mathrm{km}~\mathrm{s}^{-1}~\mathrm{Mpc}^{-1}$ (a default value in LAL adopted from Planck 2013 results~\cite{Ade:2013sjv}).

For comparison of our analysis with the results of the previous LVC analysis~\cite{Abbott:2018wiz,LIGOScientific:2018mvr}, we also analyze GW170817 by using the restricted \texttt{TF2}  approximant as the waveform model with 5+1PN-order tidal-part phase formula.
This model has the BBH baseline whose amplitude is constructed only from the Newtonian-order point-particle evolution~\cite{Buonanno:2009zt, Blanchet:2013haa, Arun:2008kb, Mikoczi:2005dn, Bohe:2013cla} and is implemented in LALINFERENCE.
We checked that 
estimates of parameters other than the tidal deformability we obtained by using the restricted \texttt{TF2} model 
are broadly consistent with
the LVC results presented in Refs.~\cite{Abbott:2018wiz,LIGOScientific:2018mvr}.

The estimates of parameters other than the tidal deformability presented in Fig.~\ref{fig:all_Low} and Table \ref{table:all_Low} 
show the absence of significant systematic difference 
associated with a difference among waveform models for both BBH baseline and tidal parts.
The posterior PDFs of these parameters for $f_\mathrm{max} = 2048~\mathrm{Hz}$ agree approximately with the ones for $f_\mathrm{max} = 1000~\mathrm{Hz}$ as illustrated for the \PNTFt~model in Fig.~\ref{fig:all_Low}. 
This is due to the fact that the parameters other than the tidal deformability 
are mainly measured from information at low frequency region~\cite{Damour:2012yf}
and terms up to 3.5PN-order of the point-particle part for the phase are the same among different waveforms.
On the other hand, the tidal deformability is measured primarily from information at high frequency region as discussed in the next section\footnote{We note that the spin-induced quadrupole moments can affect largely estimates of the component spins and mass ratio for large NSs' spins,
and combination of the effects of the spin-induced quadrupole moments and the tidal deformability
is important to investigate NS EOS as shown in Ref.~\cite{Harry:2018hke}.}.

\begin{widetext}

%----------------------------------------------------------------%
\begin{figure}[htbp]
  \begin{center}
%  \vspace{-2in}
 \begin{center}
    \includegraphics[keepaspectratio=true,height=160mm]{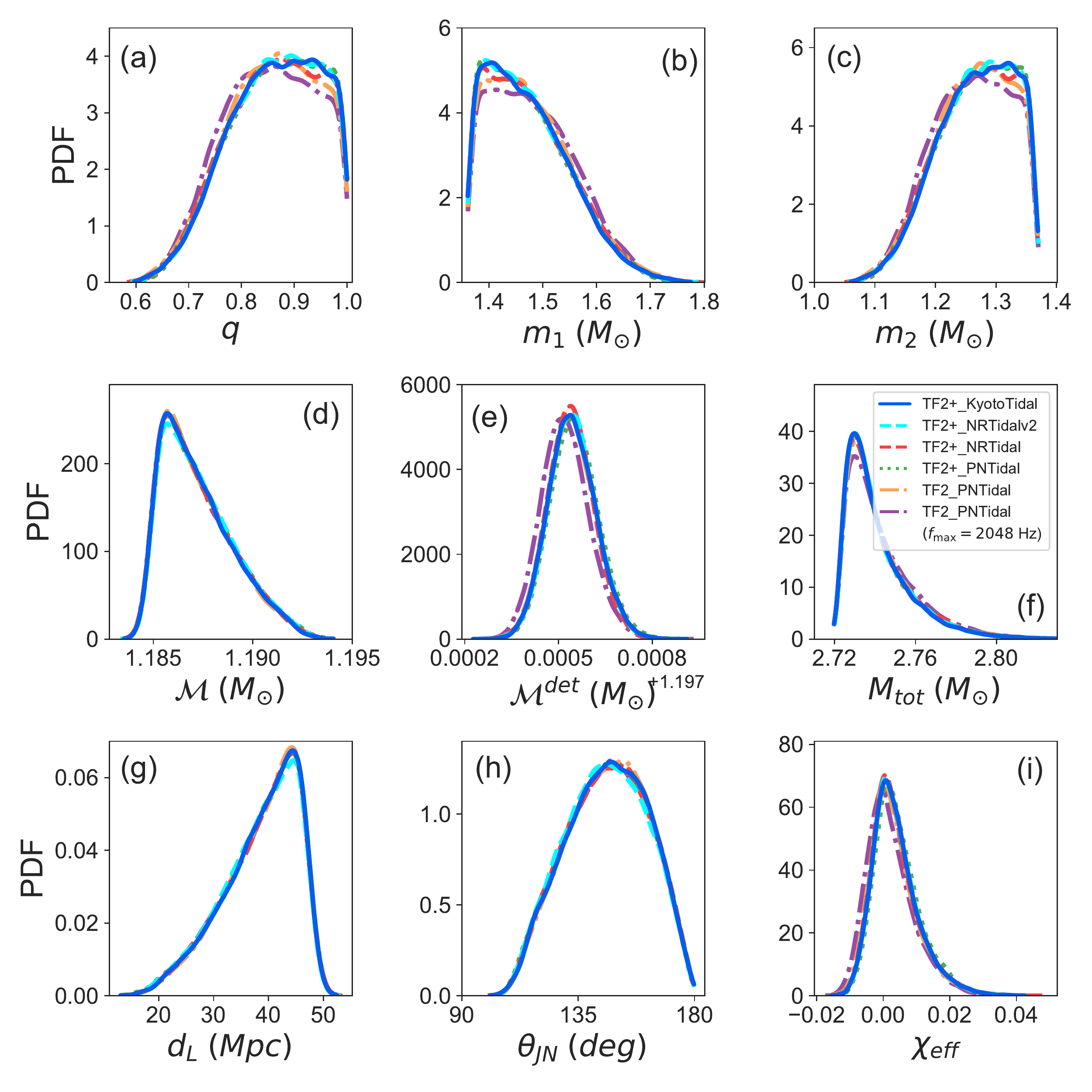}\\
 \end{center}
  \caption{
Marginalized posterior PDFs of various parameters for GW170817 derived by various waveform models.
The blue, cyan, red, green, and orange curves correspond to 
the \Kyoto, \NRTidalvt, \NRTidal, \PNTFtp, and \PNTFt~models, respectively.
The top-left, top-middle, top-right, middle-left, center, middle-right, bottom-left, bottom-middle, and bottom-right panels show (a) the mass ratio $q$, (b) the primary mass $m_1$, (c) the secondary mass $m_2$, (d) the source-frame chirp mass ${\cal M}$, (e) the detector-frame chirp mass $\mathcal{M}^\mathrm{det}$, (f) the total mass $M_\mathrm{tot}$, (g) the luminosity distance to the source $d_L$, (h) the inclination angle $\theta_\mathrm{JN}$, and (i) the effective spin parameter $\chi_\mathrm{eff}$, respectively.
Here, we show the distribution for $f_\mathrm{max} = 1000~\mathrm{Hz}$,
except for the \PNTFt~model, for which the intervals for both $f_\mathrm{max} = 1000$ and $2048~\mathrm{Hz}$ are given.
}%
\label{fig:all_Low}
\end{center}
\end{figure}
%----------------------------------------------------------------%

\begin{table}[htbp]
\begin{center}
\begin{tabular}{lccccc}
\hline \hline
 & ~~\PNTFt & ~~\PNTFtp &  ~~\Kyoto &  ~~\NRTidal &  ~~\NRTidalvt \\ \hline
Luminosity distance $d_{L}$ [Mpc] & $40.0^{+7.3}_{-14.4}$ & $39.8^{+7.5}_{-14.7}$ & $39.9^{+7.3}_{-14.6}$ & $39.9^{+7.4}_{-14.5}$ & $39.6^{+7.7}_{-14.6}$ \\
Binary inclination $\theta_\mathrm{JN}$ [degree] & $147^{+24}_{-32}$ & $146^{+24}_{-27}$ & $147^{+24}_{-28}$ & $147^{+24}_{-27}$ & $146^{+25}_{-27}$ \\
Detector-frame chirp mass ${\cal M}^\mathrm{det}~[\solM]$ & $1.1975^{+0.0001}_{-0.0001}$ & $1.1975^{+0.0001}_{-0.0001}$ & $1.1975^{+0.0001}_{-0.0001}$ & $1.1975^{+0.0001}_{-0.0001}$ & $1.1975^{+0.0001}_{-0.0001}$ \\
Source-frame chirp mass ${\cal M}~[\solM]$ & $1.187^{+0.004}_{-0.002}$ & $1.187^{+0.004}_{-0.002}$ & $1.187^{+0.004}_{-0.002}$ & $1.187^{+0.004}_{-0.002}$ & $1.187^{+0.004}_{-0.002}$ \\
Primary mass $m_1~[\solM]$ & $(1.36,~1.59)$ & $(1.36,~1.58)$ & $(1.36,~1.58)$ & $(1.36,~1.59)$ & $(1.36,~1.58)$ \\   
Secondary mass $m_2~[\solM]$ & $(1.18,~1.37)$ & $(1.18,~1.37)$ & $(1.18,~1.37)$ & $(1.18,~1.37)$ & $(1.18,~1.37)$ \\
Total mass $M_\mathrm{tot}:=m_1+m_2~[\solM]$ & $2.74^{+0.04}_{-0.01}$ & $2.74^{+0.04}_{-0.01}$ & $2.74^{+0.04}_{-0.01}$ & $2.74^{+0.04}_{-0.01}$ & $2.74^{+0.04}_{-0.01}$ \\
Mass ratio $q:=m_2/m_1$ & $(0.74,~1.00)$ & $(0.74,~1.00)$ & $(0.75,~1.00)$ & $(0.75,~1.00)$ & $(0.75,~1)$ \\
Effective spin $\chi_\mathrm{eff}$ & $0.002^{+0.015}_{-0.009}$ & $0.003^{+0.015}_{-0.009}$ & $0.003^{+0.014}_{-0.008}$ & $0.002^{+0.015}_{-0.008}$ & $0.003^{+0.014}_{-0.008}$ \\
\hline \hline
\end{tabular}
\caption{
90\% credible interval of the luminosity distance $d_L$, the binary inclination $\theta_\mathrm{JN}$, mass parameters,
and the effective spin parameter $\chi_\mathrm{eff}$ for GW170817 estimated using various waveform models. 
We show 10\%--100\% regions of the mass ratio with the upper limit $q = 1$ imposed by definition, 
and those of $m_1$ and $m_2$ are given accordingly. 
We give symmetric 90\% credible intervals, i.e., 5\%--95\%, for the other parameters 
with the median as a representative value. 
}
\label{table:all_Low}
\end{center}
\end{table}

%----------------------------------------------------------------%
\begin{figure}[htbp]
  \begin{center}
\begin{tabular}{cc}
 \begin{minipage}[b]{0.45\linewidth}
 \begin{center}
    \includegraphics[keepaspectratio=true,height=80mm]{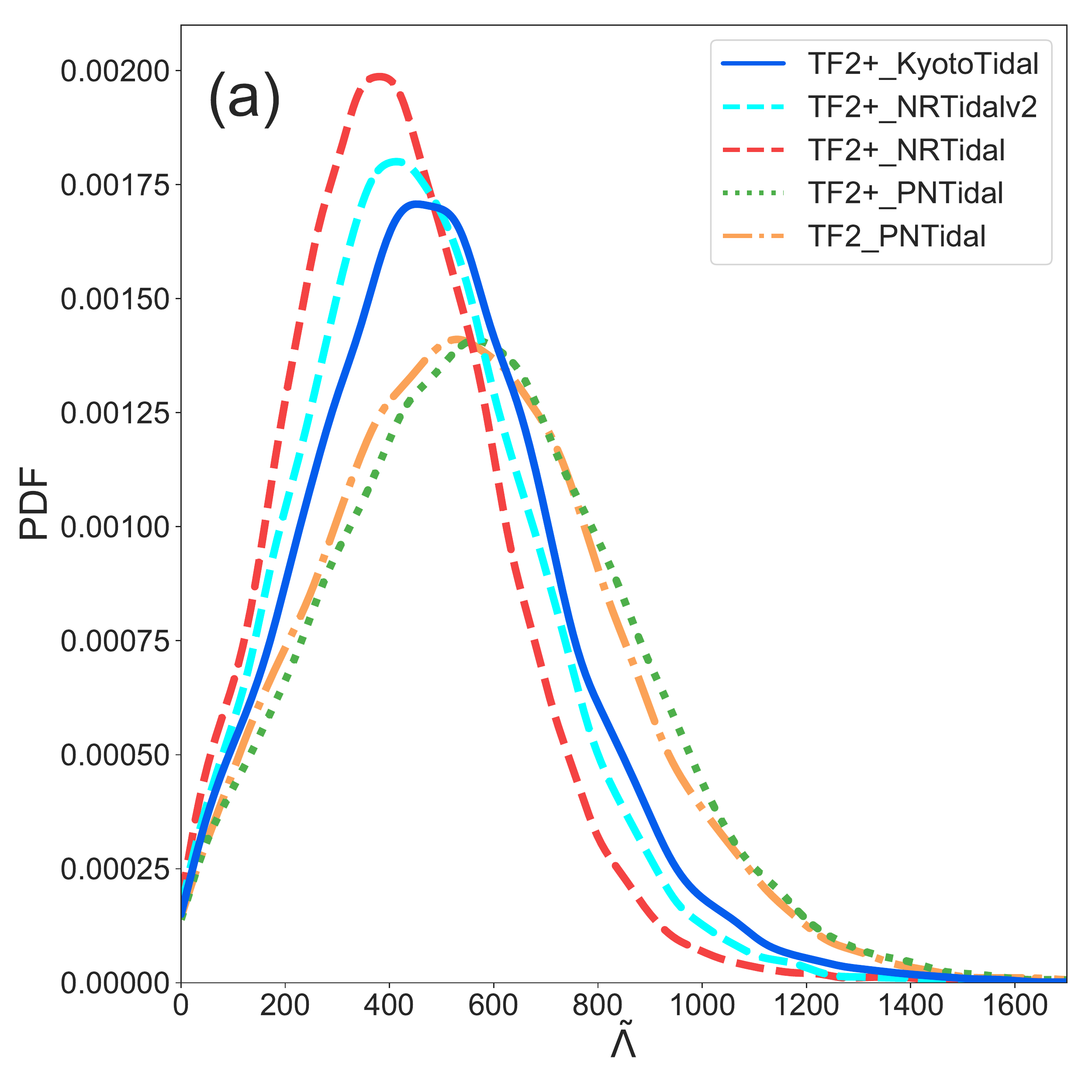}\\
 \end{center}
 \end{minipage}
 \begin{minipage}[b]{0.45\linewidth}
  \begin{center}
    \includegraphics[keepaspectratio=true,height=80mm]{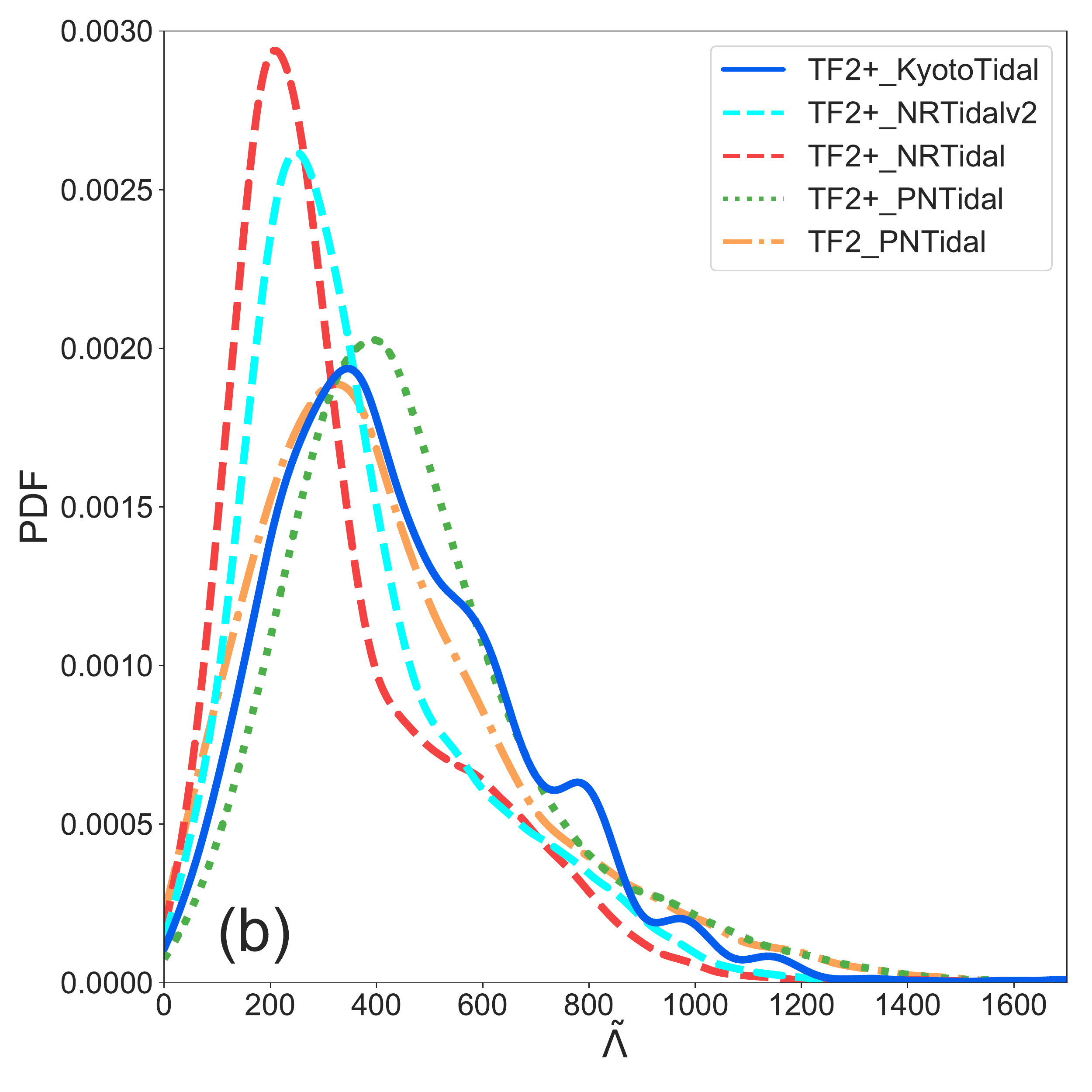}\\
 \end{center}
 \end{minipage}
\end{tabular}
  \caption{
Marginalized posterior PDFs of binary tidal deformability, $\tilde{\Lambda}$, for GW170817,
estimated by various waveform models, for (a) $f_\mathrm{max} = 1000~\mathrm{Hz}$ and (b) 2048 Hz.
The blue, cyan, red, green, and orange curves correspond to 
the \Kyoto, \NRTidalvt, \NRTidal, \PNTFtp, and \PNTFt~models, respectively.
The corresponding 90\% credible intervals are presented in Table \ref{table:lamt}.
}%
\label{fig:lamt}
\end{center}
\end{figure}
%----------------------------------------------------------------%

\begin{table}[htbp]
\begin{center}
\begin{tabular}{l|cc|cc}
\hline \hline
Model && $f_\mathrm{max} = 1000~\mathrm{Hz}$ &&  $f_\mathrm{max} = 2048~\mathrm{Hz}$\\ \hline
           & Symmetric & HPD & Symmetric & HPD \\ \hline
\PNTFt & $548^{+500}_{-415}$ & $548^{+433}_{-463}$ & $376^{+584}_{-284}$ & $376^{+442}_{-353}$ \\ 
\PNTFtp & $569^{+496}_{-431}$ & $569^{+441}_{-470}$ & $428^{+540}_{-280}$ & $428^{+414}_{-353}$ \\ 
\Kyoto & $481^{+436}_{-359}$ & $481^{+379}_{-402}$ & $402^{+465}_{-279}$ & $402^{+419}_{-316}$ \\ 
\NRTidal & $403^{+378}_{-299}$ & $403^{+328}_{-337}$ & $267^{+491}_{-180}$ & $267^{+409}_{-228}$ \\ 
\NRTidalvt & $445^{+412}_{-330}$ & $445^{+357}_{-370}$ & $312^{+498}_{-208}$ & $312^{+407}_{-263}$ \\ 
\hline \hline
\end{tabular}
\caption{
90\% credible interval of the binary tidal deformability, $\tilde{\Lambda}$, for GW170817 for various waveform models.
We report both the symmetric 90\% credible interval (Symmetric) and the 90\% highest-posterior-density (HPD) intervals, 
for both $f_\mathrm{max} = 1000~\mathrm{Hz}$ (left side) and 2048 Hz (right side), where the median is shown as a representative value.
}
\label{table:lamt}
\end{center}
\end{table}

\end{widetext}

%%%%%%%%%%%%%%%%%%%%%%%%%%%%%%%%%%%%%%%%%%%%%%%
\subsection{Posterior of binary tidal deformability}
\label{sec:Lambdatilde}
%%%%%%%%%%%%%%%%%%%%%%%%%%%%%%%%%%%%%%%%%%%%%%%
Before presenting our results obtained with various waveform models, we first compare our results obtained by using the restricted \texttt{TF2} model that incorporates the 5+1PN-order tidal-part phase with those from the LVC analysis~\cite{Abbott:2018wiz} as a sanity check. 
The restricted \texttt{TF2} model version used by LVC analysis includes no amplitude 
corrections and has a uniform prior on the component tidal deformability, with 
$\Lambda_{1,2}\sim U[0,~5000]$. 
While our result of 90\% credible symmetric and highest posterior density (HPD) intervals on $\tilde{\Lambda}$ are 
$347^{+564}_{-243}$ and $347^{+453}_{-295}$, respectively, for restricted \texttt{TF2} with 5+1PN-order tidal-part phase, low-spin prior ($|\chi_{1,2}| \leq 0.05$), and $f_\mathrm{max} = 2048~\mathrm{Hz}$,
the LVC report $\tilde{\Lambda} = 340^{+580}_{-240}$ and $340^{+490}_{-290}$, respectively, in \cite{Abbott:2018wiz}.
Here, uniform priors in $\Lambda_1$ and $\Lambda_2$ are adopted in both analyses,
and the posterior of $\tilde{\Lambda}$ is divided by its prior determined by those of other parameters 
following \cite{Abbott:2018wiz}
to derive approximate results for the case of a uniform prior on $\tilde{\Lambda}$.
The closeness of the inferred credible ranges indicates that our analysis successfully reproduces the results derived by the LVC.
If we assume a uniform prior on $\tilde{\Lambda}$ from the beginning, 90\% credible symmetric and HPD intervals on $\tilde{\Lambda}$ are $316^{+504}_{-224}$ and $316^{+367}_{-291}$, respectively, for restricted \texttt{TF2} with 5+1PN-order tidal-part phase.

Figure~\ref{fig:lamt} shows the marginalized posterior PDFs
for the binary tidal deformability $\tilde{\Lambda}$ for various waveform models with both (a) $f_\mathrm{max} = 1000~\mathrm{Hz}$ and (b) 2048 Hz.
The corresponding 90\% credible intervals are presented in Table~\ref{table:lamt}.
We caution that the \Kyoto~model is calibrated only up to 1000 Hz and can overestimate tidal effects
at frequencies above $1000~\mathrm{Hz}$.
Thus, 
the results for $f_\mathrm{max} = 2048~\mathrm{Hz}$ should be regarded as only a reference.

For $f_\mathrm{max} = 1000~\mathrm{Hz}$ [see Fig.~\ref{fig:lamt} (a)], 
the peak values of $\tilde{\Lambda}$ are located between 400 and 500 and the 90\% credible intervals do not extend to $\gtrsim900$
for NR calibrated waveform models: the \Kyoto, \NRTidalvt, and \NRTidal~models.
Our results show that the posterior of binary tidal deformability for GW170817 
depends on waveform models.
The \Kyoto, \NRTidal, \NRTidalvt, and \PNTFtp~models are constructed from the same BBH baseline, \TFtp, 
but with different tidal descriptions.
Therefore, a difference of estimates among these waveform models reflects directly their different tidal description.
The \NRTidal~model gives the smallest median value on $\tilde{\Lambda}$ of 403, 
the second is the \NRTidalvt~model of 445,
the third is the \Kyoto~model of 481, 
and the \PNTFtp~model gives the largest one of 569.
This order is derived from the order of the phase shift of different waveform models
for a given value of $\tilde{\Lambda}=400$, up to about 1400 Hz as shown in Fig.~\ref{fig:TidalWFs}.
The tendency to give smaller estimated values for NR calibrated waveform models than for PN waveform models
are consistent with previous results derived in Ref.~\cite{Samajdar:2018dcx} (see also Ref.~\cite{Samajdar:2019ulq} for the detailed study of systematic biases associated with spin effects).
The \PNTFtp~and \PNTFt~models 
are constructed from the same tidal part and the different point-particle part.
A difference in the posterior PDFs of estimated $\tilde{\Lambda}$ between these models is very small for $f_\mathrm{max}=1000~\mathrm{Hz}$.
This result shows that the higher-order point-particle terms do not significantly affect the estimate of the binary tidal deformability of GW170817 for $f_\mathrm{max}=1000~\mathrm{Hz}$.
(See Ref. \cite{Messina:2019uby} for systematic study on the binary tidal deformability 
by injection of signals with incomplete baselines)

For $f_\mathrm{max} = 2048~\mathrm{Hz}$ [see Fig.~\ref{fig:lamt} (b)],
the peak values of $\tilde{\Lambda}$ are located between 250 and 400 and the 90\% credible intervals do not extend to $\gtrsim850$
for NR calibrated waveform models.
The widths of symmetric 90\% credible intervals for $f_\mathrm{max} = 2048~\mathrm{Hz}$
are narrower than those for $f_\mathrm{max} = 1000~\mathrm{Hz}$, 
by about 7\% for the \Kyoto~model, 
4\% for the \NRTidal~model, 
5\% for the \NRTidalvt~model,
13\% for the \PNTFtp~model, 
and about 5\% for the \PNTFt~model, as shown in Table~\ref{table:lamt}.
This decrease in the width of the interval is consistent with the fact that higher-frequency data are more informative to measure $\tilde{\Lambda}$~\cite{Damour:2012yf}.
The peak values of the posterior PDFs of $\tilde{\Lambda}$ tend to decrease as $f_\mathrm{max}$ increases for all waveform models as shown in Fig.~\ref{fig:lamt}.
The order of peak values of $\tilde{\Lambda}$ for the different waveform models 
that incorporate the same BBH baseline, \texttt{TF2+},
is not affected by varying $f_\mathrm{max}$
as shown in Fig.~\ref{fig:lamt}.
This is explained by the same reason as that for $f_\mathrm{max}=1000~\mathrm{Hz}$.
We note that 1400 Hz approximately corresponds to $f_\mathrm{ISCO}$ for estimated mass range.
The \PNTFt~model gives a slightly smaller peak value than the \Kyoto~model.
This cannot be explained only by the feature of the tidal part as shown in Fig.~\ref{fig:TidalWFs}.
This might be due to the effects of the higher-order point-particle terms 
or the fact that the data at frequencies above $1000~\mathrm{Hz}$ are dominated by the detector's noise.
The difference in the posterior PDFs of estimated $\tilde{\Lambda}$ between the \PNTFtp~and \PNTFt~models for $f_\mathrm{max}=2048~\mathrm{Hz}$
is larger than that for $f_\mathrm{max}=1000~\mathrm{Hz}$ (see Fig.~\ref{fig:lamt} and Table \ref{table:lamt}). This is due to the effects of higher-order point-particle terms as discussed in Refs.~\cite{Yagi:2013baa,Messina:2019uby}.

%%%%%%%%%%%%%%%%%%%%%%%%%%%%%%%%%%%%%%%%%%%%%%%
\section{Results of GW190425 and NS EOS}
\label{sec:GW190425_EOS}
%%%%%%%%%%%%%%%%%%%%%%%%%%%%%%%%%%%%%%%%%%%%%%%
We reanalyze data of the second BNS merger event, GW190425, 
using three waveform models; \Kyoto, \NRTidalvt, and \PNTFtp~models. 
We present marginalized posterior PDFs for source parameters of GW190425 in Fig. \ref{fig:all_Low:GW190425}
and corresponding 90\% credible interval in Table \ref{table:all_Low:GW190425}.
The estimates of parameters other than $\tilde{\Lambda}$ presented in Fig. \ref{fig:all_Low:GW190425}
and Table \ref{table:all_Low:GW190425} 
are broadly consistent with the LVC results presented in Ref.~\cite{Abbott:2020uma}
and show the absence of significant systematic difference among different waveform models.
The posterior PDFs of these parameters for $f_\mathrm{max} = 2048~\mathrm{Hz}$ agree approximately with the ones for $f_\mathrm{max} = 1000~\mathrm{Hz}$ as illustrated for the \PNTFtp~model in Fig.~\ref{fig:all_Low:GW190425}.

Figure \ref{fig:lamt:GW190425} shows marginalized posterior PDFs of $\tilde{\Lambda}$ for GW190425
with three waveform models.
While this figure indicates that there is a small difference between PN and NR calibrated models,
only a tiny difference is found between two NR calibrated models.
The posterior PDF of $\tilde{\Lambda}$ has a large value around $\tilde{\Lambda}=0$
and this fact implies that no significant tidal effect is detected as found in Ref.~\cite{Abbott:2020uma}.
HPD upper limit on the binary tidal deformability
is $\tilde{\Lambda}\leq610$ for the \Kyoto~model for $f_\mathrm{max}=1000~\mathrm{Hz}$.
The posterior PDF of $\tilde{\Lambda}$ for the \Kyoto~model with $f_\mathrm{max}=2048~\mathrm{Hz}$ is bimodal.
Investigation of the secondary peak's origin remains as a future work,
but it may result from the nonlinear tidal terms $\propto x^p$ of this model, which increase rapidly at $\gtrsim 1000$ Hz for which the calibration by the hybrid waveforms is not performed.

%%%%%% GW190425 %%%%%

\begin{widetext}

%----------------------------------------------------------------%
\begin{figure}[htbp]
  \begin{center}
 \begin{center}
    \includegraphics[keepaspectratio=true,height=160mm]{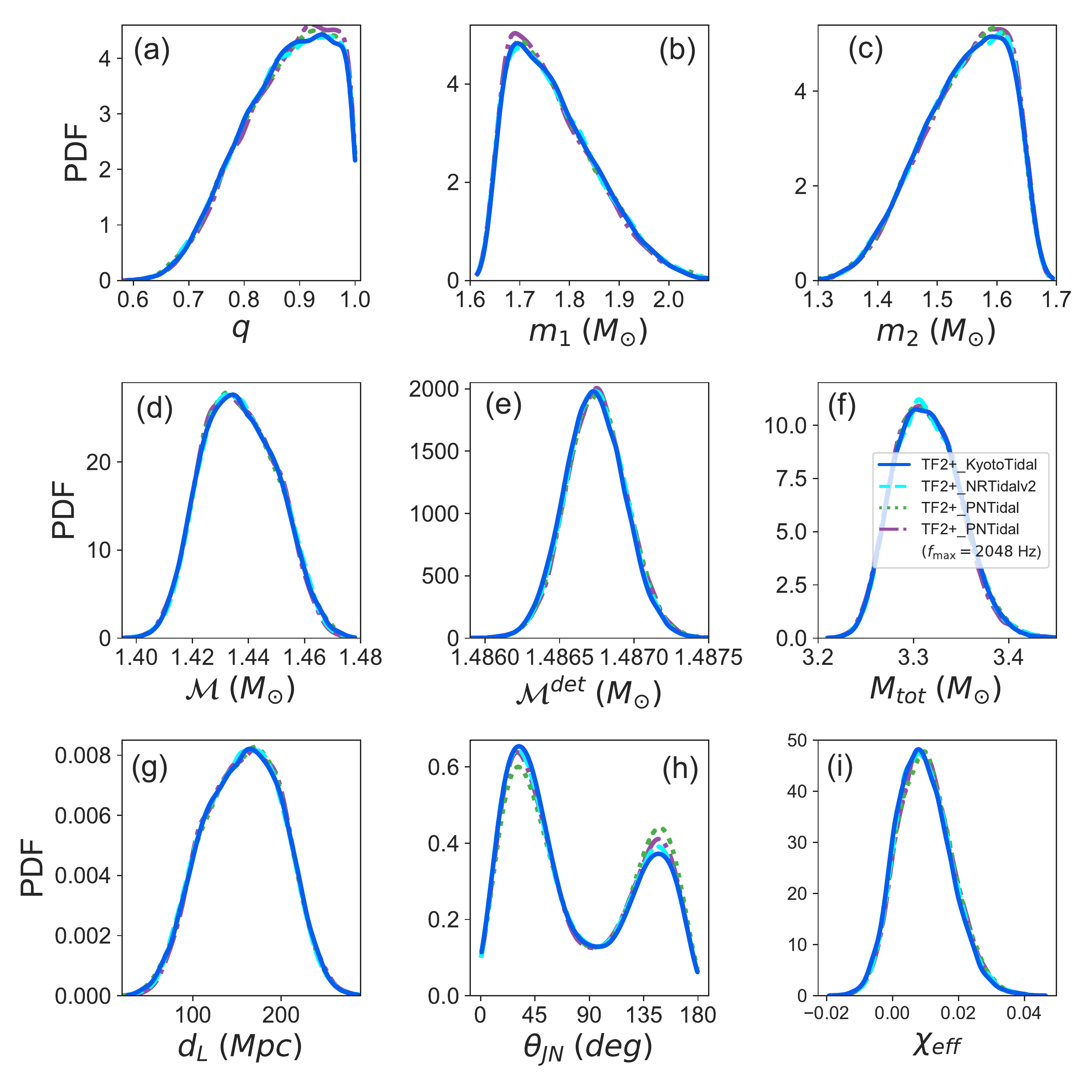}\\
 \end{center}
  \caption{
Marginalized posterior PDFs of source parameters for GW190425 
using \PNTFtp, \Kyoto, and \NRTidalvt~models
for the low-spin prior ($|\chi_{1,2}|\leq0.05$).
Here, we show the distribution for $f_\mathrm{max}=1000~\mathrm{Hz}$,
except for the \PNTFtp~model, for which the intervals for both $f_\mathrm{max}=1000$ and $2048~\mathrm{Hz}$ are given.
}%
\label{fig:all_Low:GW190425}
\end{center}
\end{figure}
%----------------------------------------------------------------%

\begin{table}[htbp]
\begin{center}
\begin{tabular}{lccccc}
\hline \hline
 & ~~\PNTFtp &  ~~\Kyoto &  ~~\NRTidalvt \\ \hline
Luminosity distance $d_{L}$ [Mpc] & $159^{+67}_{-74}$ & $159^{+67}_{-73}$ & $158^{+67}_{-73}$ \\
Detector-frame chirp mass ${\cal M}^\mathrm{det}~[\solM]$ & $1.4867^{+0.0003}_{-0.0003}$ & $1.4867^{+0.0003}_{-0.0003}$ & $1.4867^{+0.0003}_{-0.0003}$ \\
Source-frame chirp mass ${\cal M}~[\solM]$ & $1.44^{+0.02}_{-0.02}$ & $1.44^{+0.02}_{-0.02}$ & $1.44^{+0.02}_{-0.02}$ \\
Primary mass $m_1~[\solM]$ & $(1.62,~1.90)$ & $(1.61,~1.90)$ & $(1.61,~1.90)$ \\   
Secondary mass $m_2~\solM$ & $(1.44,~1.69)$ & $(1.44,~1.69)$ & $(1.44,~1.69)$ \\
Total mass $M_\mathrm{tot}:=m_1+m_2~[\solM]$ & $3.3^{+0.1}_{-0.1}$ & $3.3^{+0.1}_{-0.1}$ & $3.3^{+0.1}_{-0.1}$ \\
Mass ratio $q:=m_2/m_1$ & $(0.8,~1.0)$ & $(0.8,~1.0)$ & $(0.8,~1.0)$ \\
Effective spin $\chi_\mathrm{eff}$ & $0.010^{+0.015}_{-0.012}$ & $0.009^{+0.015}_{-0.012}$ & $0.009^{+0.015}_{-0.012}$ \\
Binary tidal deformability $\tilde{\Lambda}$ & $\le700$ & $\le610$ & $\le546$  \\
\hline \hline
\end{tabular}
\caption{
Source properties for GW190425
using \PNTFtp, \Kyoto, and \NRTidalvt~models
for $f_\mathrm{max}=1000~\mathrm{Hz}$
for the low-spin prior ($|\chi_{1,2}|\leq0.05$).
For $\tilde{\Lambda}$, we show HPD upper limits.
}
\label{table:all_Low:GW190425}
\end{center}
\end{table}

%----------------------------------------------------------------%
\begin{figure}[htbp]
  \begin{center}
\begin{tabular}{cc}
 \begin{minipage}[b]{0.45\linewidth}
 \begin{center}
    \includegraphics[keepaspectratio=true,height=80mm]{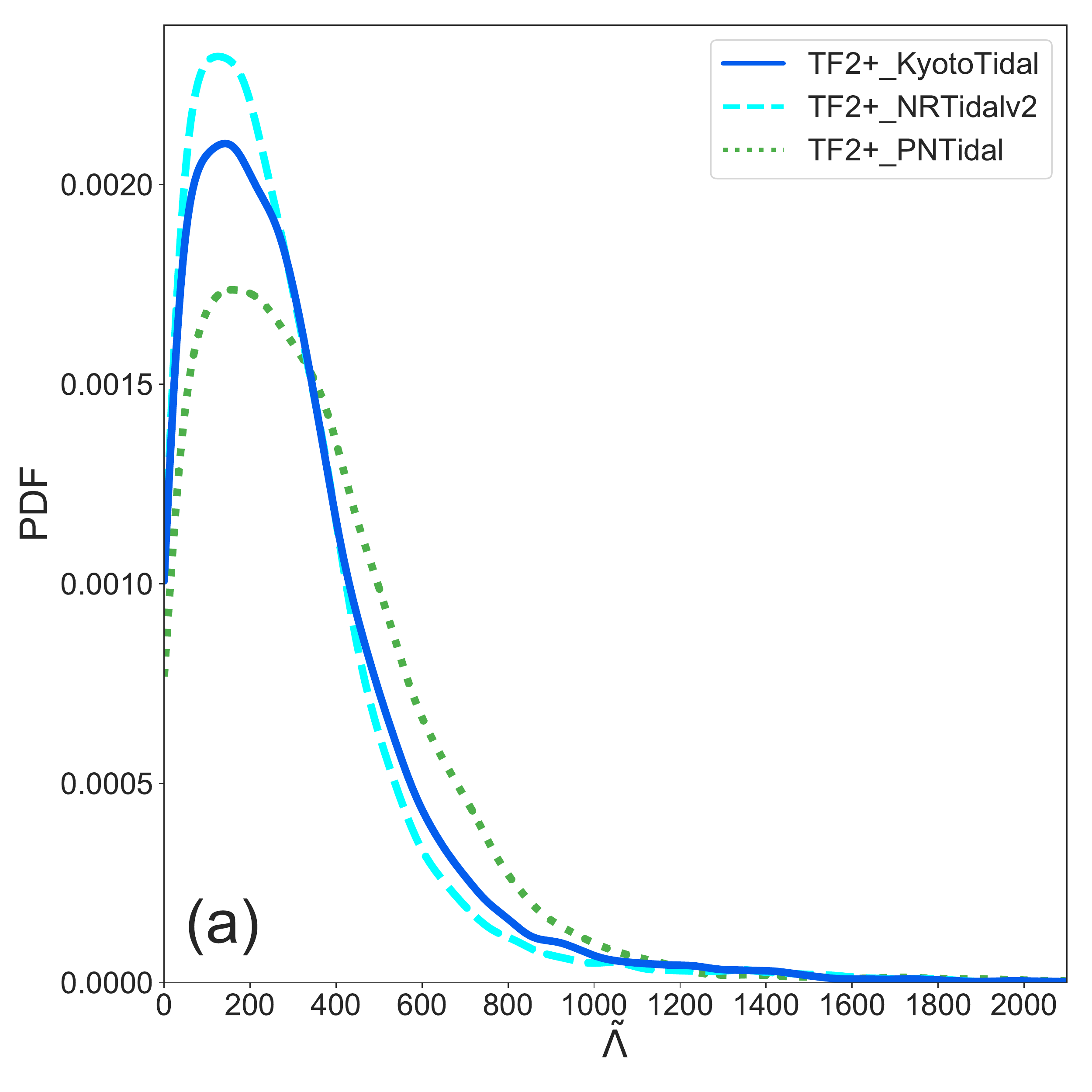}\\
 \end{center}
 \end{minipage}
 \begin{minipage}[b]{0.45\linewidth}
  \begin{center}
    \includegraphics[keepaspectratio=true,height=80mm]{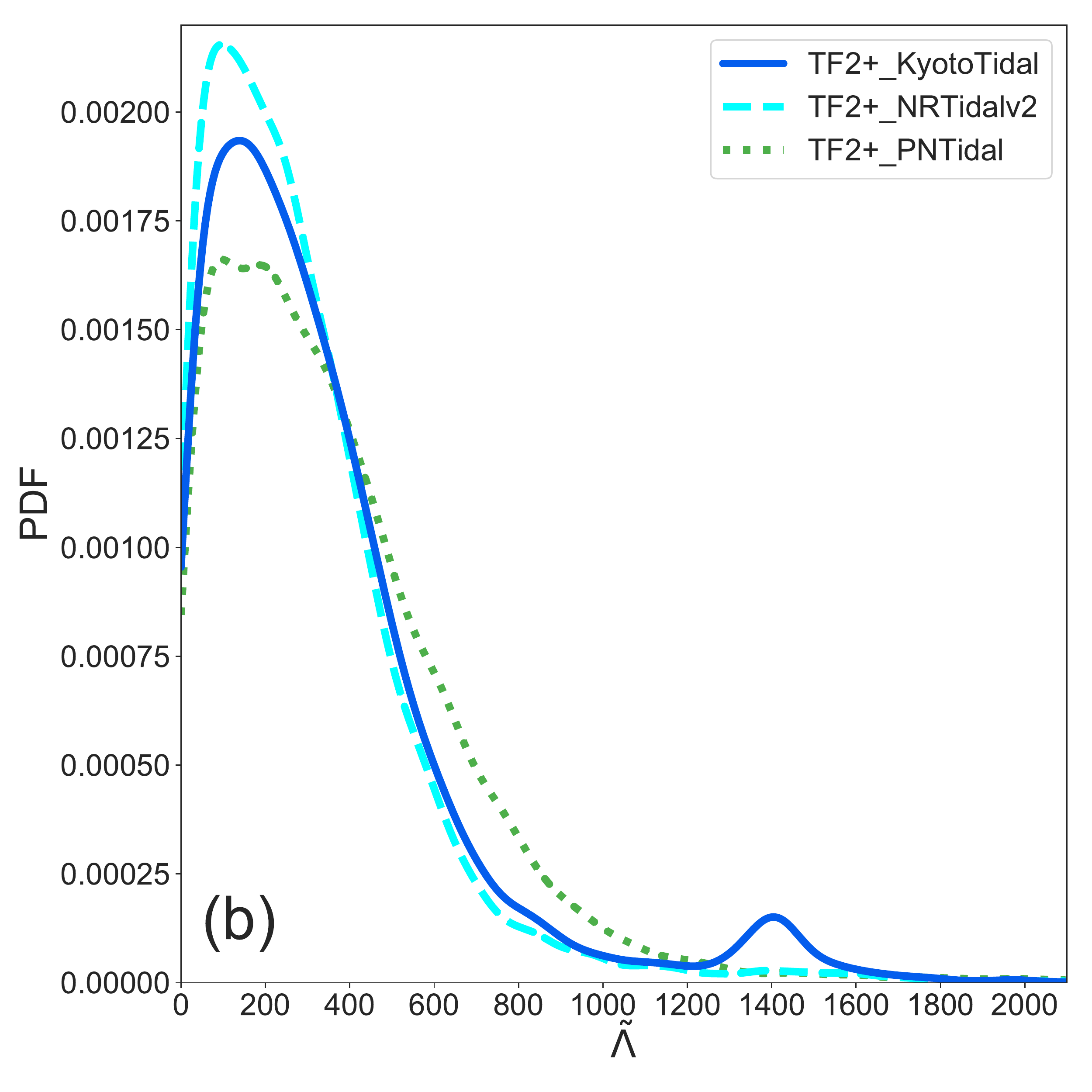}\\
 \end{center}
 \end{minipage}
\end{tabular}
  \caption{
Marginalized posterior PDFs of $\tilde{\Lambda}$ for GW190425
using \PNTFtp, \Kyoto, and \NRTidalvt~models for the low-spin prior ($|\chi_{1,2}|\leq0.05$)
for (a) $f_\mathrm{max}=1000$ and (b) $2048~\mathrm{Hz}$.
}%
\label{fig:lamt:GW190425}
\end{center}
\end{figure}
%----------------------------------------------------------------%

%----------------------------------------------------------------%
\begin{figure}[htbp]
  \begin{center}
\begin{tabular}{cc}
 \begin{minipage}[b]{0.45\linewidth}
 \begin{center}
    \includegraphics[keepaspectratio=true,height=80mm]{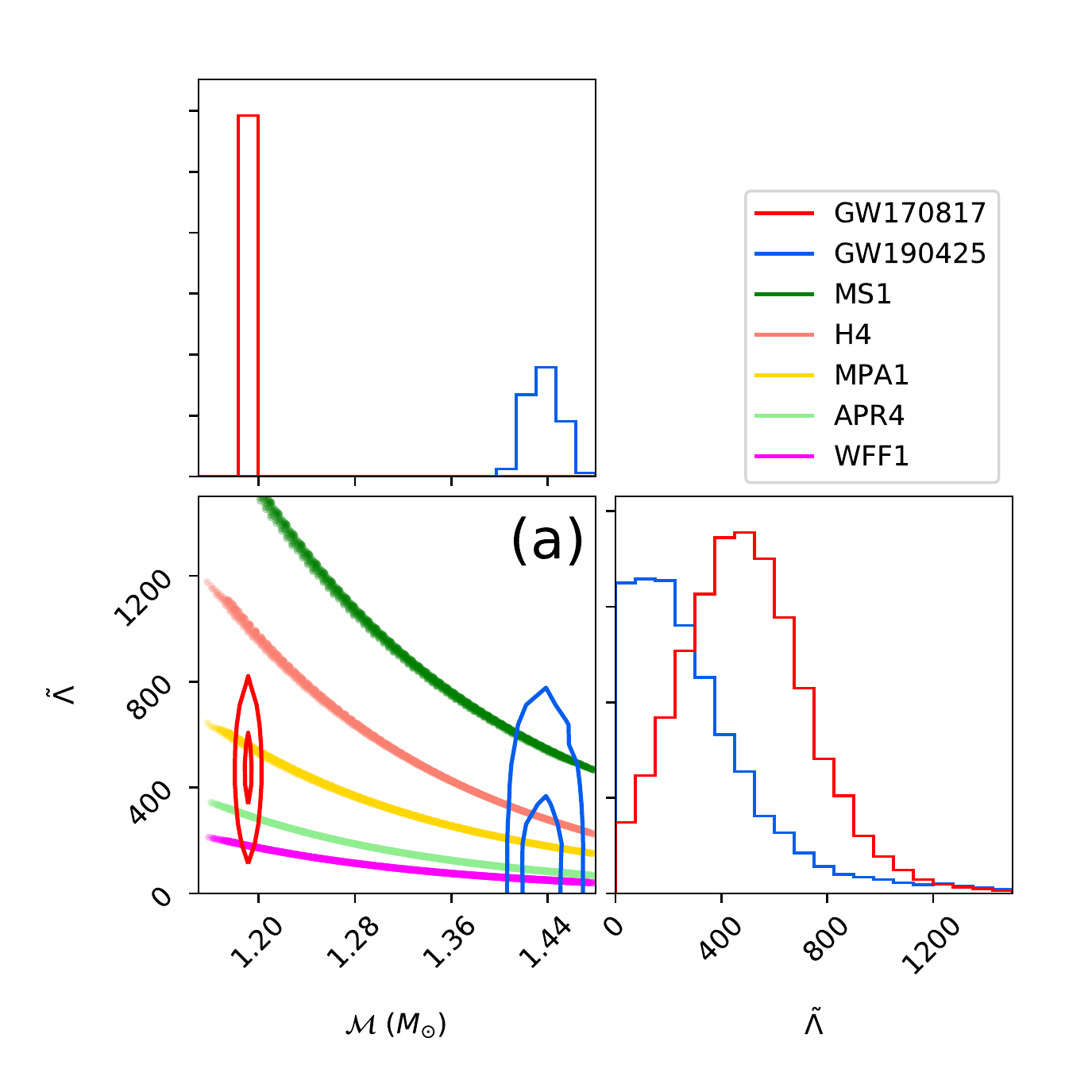}\\
 \end{center}
 \end{minipage}
 \begin{minipage}[b]{0.45\linewidth}
  \begin{center}
    \includegraphics[keepaspectratio=true,height=80mm]{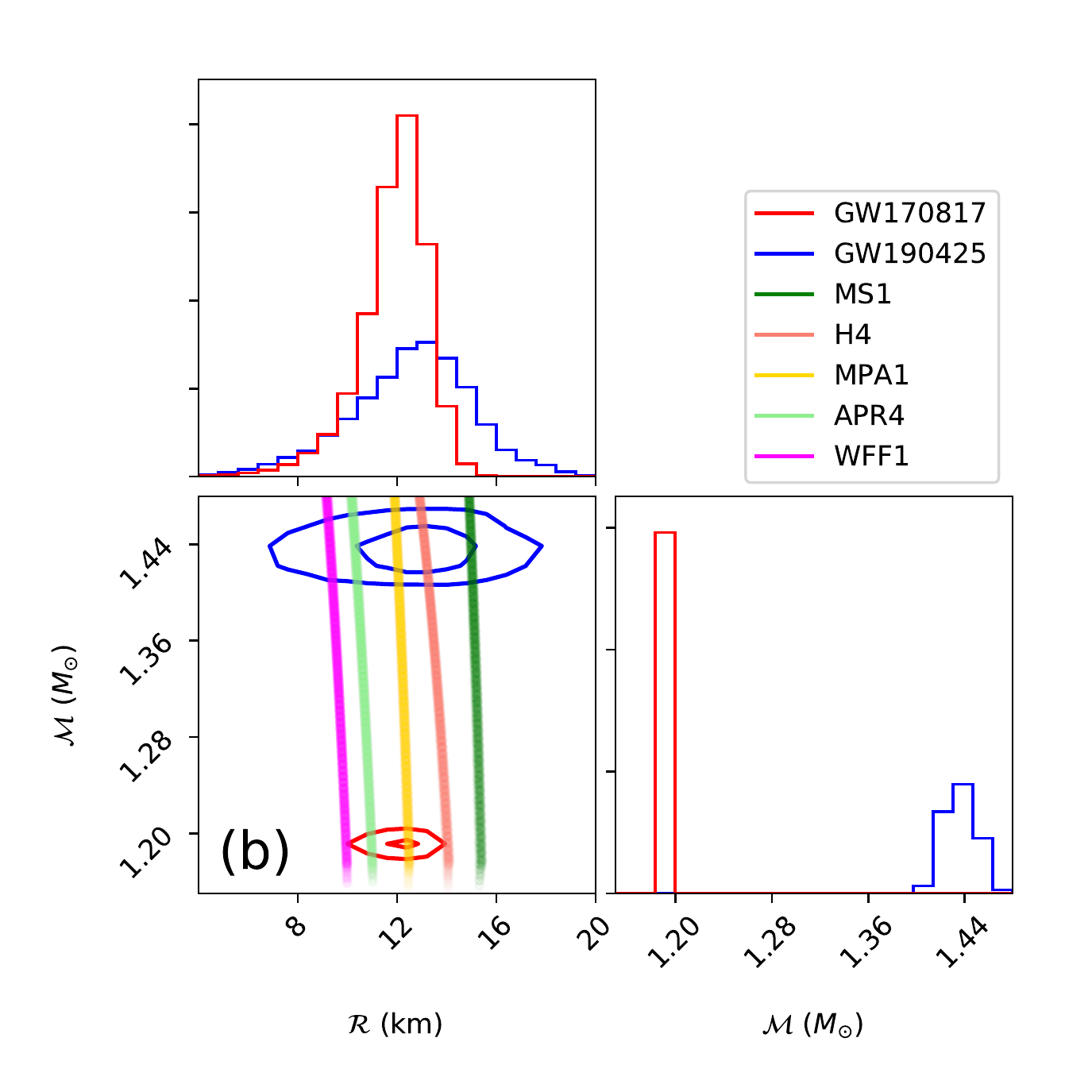}\\
 \end{center}
 \end{minipage}
\end{tabular}
  \caption{
50\% and 90\% credible regions in (a) the $\tilde{\Lambda}$-${\cal M}$ plane  
and (b) the ${\cal M}$-${\cal R}$ plane 
obtained from GW170817 (red) and GW190425 (blue) 
events using \Kyoto~model for $f_\mathrm{max}=1000~\mathrm{Hz}$
for the low-spin prior ($|\chi_{1,2}|\leq0.05$).
Five colored curves are calculated in the mass ratio range $0.7\leq q \leq 1$ 
with various EOS models; MS1, H4, MPA1, APR4, and WFF1.
}%
\label{fig:Lamt-Mc_Mc-Rc}
\end{center}
\end{figure}
%----------------------------------------------------------------%
\end{widetext}

%%%%%%%%%%%%%%%%%%%%%%%%%%%%%%%%%%%%%%%%%%%%%%%
%\section{Constraints on NS EOS models with GW170817 and GW190425}
%%%%%%%%%%%%%%%%%%%%%%%%%%%%%%%%%%%%%%%%%%%%%%%
In order to discuss constraints on NS EOS by combining information obtained from GW170817 and GW190425,
we plot prediction of various NS EOS on posterior of the binary tidal deformability and binary's chirp radius, 
which is a conveniently scaled dimensionful radius-like parameter \cite{Wade:2014vqa}.
Figure \ref{fig:Lamt-Mc_Mc-Rc} shows 
50\% and 90\% credible regions in (a) the $\tilde{\Lambda}$-${\cal M}$ plane 
and (b) the ${\cal M}$-${\cal R}$ plane, for GW170817 and GW190425,
where ${\cal R}=2{\cal M}\tilde{\Lambda}^{1/5}$ is the binary's chirp radius. 
Five colored curves are posteriors predicted by various NS EOS models;
MS1 \cite{Mueller:1996pm}, H4 \cite{Lackey:2005tk}, MPA1 \cite{Muther:1987xaa}, 
APR4 \cite{Akmal:1998cf}, and WFF1 \cite{Wiringa:1988tp}.
For these plots, we use the masses uniformly distributed in 
the mass ratio range $0.7\leq q \leq1$,
which include the 90\% credible regions of mass posteriors for GW170817 and GW190425. 
Our results using \Kyoto~model show that the allowed chirp radius range is $10~\mathrm{km}\lesssim\mathcal{R}\lesssim14~\mathrm{km}$ for the chirp mass $\mathcal{M}\simeq1.2~M_\odot$.
In particular, the MS1 and H4 models lie outside the 90\% credible region for GW170817,
while they are not disfavored from GW190425, which is consistent with the LVC results presented in Refs.~\cite{TheLIGOScientific:2017qsa,Abbott:2018wiz,Abbott:2018exr,LIGOScientific:2019eut}.

%%%%%%%%%%%%%%%%%%%%%%%%%%%%%%%%%%%%%%%%%%%%%%%
\section{Summary}
\label{sec:summary}
%%%%%%%%%%%%%%%%%%%%%%%%%%%%%%%%%%%%%%%%%%%%%%%
We reanalyze GW170817 and GW190425 with a NR calibrated waveform model, the \Kyoto~model, which has been developed independently from the one used in previous studies by LVC.
The \Kyoto~model is calibrated in the frequency range of 10--1000 Hz to hybrid waveforms composed of 
high-precision NR waveforms and the \texttt{SEOBNRv2T} waveforms, and reproduces the phase of the hybrid waveforms within 0.1 rad error up to $1000~\mathrm{Hz}$.
In the \Kyoto~model, the nonlinear effects of the tidal deformability are incorporated.
We also reanalyze the events with other waveform models: two PN waveform models (\PNTFt~and \PNTFtp), the \NRTidal~model that is another NR calibrated waveform model, and its upgrade, the \NRTidalvt~model.

We compare parameter estimation results with those by different tidal waveform models.
For GW170817, we do not find any significant systematic differences for extraction of source parameters other than the binary tidal deformability 
using different waveform models.
By contrast, we find the significant systematics in determining $\tilde{\Lambda}$. 
Specifically, we reconfirm that the PN model tends to overestimate $\tilde{\Lambda}$ 
compared to the NR calibrated waveform models as shown in Ref.~\cite{Samajdar:2018dcx},
and in addition, the estimates of $\tilde{\Lambda}$ depend on NR calibrated waveform models
for $f_\mathrm{max}=1000~\mathrm{Hz}$
although the difference is smaller than the statistical uncertainties.

Our results for GW170817 indeed indicate that 
$\tilde{\Lambda}$ is constrained more tightly for $f_\mathrm{max} = 2048~\mathrm{Hz}$ than for $f_\mathrm{max} = 1000~\mathrm{Hz}$.
For the \Kyoto~model, the 90\% symmetric interval of $\tilde{\Lambda}$ for $f_\mathrm{max} = 2048~\mathrm{Hz}$ is about 7\% narrower than that for $f_\mathrm{max} = 1000~\mathrm{Hz}$.
Although the credible interval of $\tilde{\Lambda}$ becomes narrower as the $f_\mathrm{max}$ increases,
the \Kyoto~model is calibrated only up to $1000~\mathrm{Hz}$.
Since higher frequency data are more informative for $\tilde{\Lambda}$~\cite{Damour:2012yf},
it is important to improve current waveform models 
at high-frequencies above $1000~\mathrm{Hz}$
to accurately determine $\tilde{\Lambda}$ from the GW data, toward third generation detector era.

For the second BNS merger event, GW190425, we use three waveform models;
\Kyoto, \NRTidalvt, and \PNTFtp~models.
Similarly to GW170817, we do not find any significant systematic differences
for extraction of source parameters other than the binary tidal deformability 
among different waveform models.
This binary system is massive and it is intrinsically difficult to measure the tidal effect.
While our results show that the 90\% credible interval of $\tilde{\Lambda}$ for the PN waveform model 
is slightly wider than for NR models, 
we do not find essential difference in the constraints for $\tilde{\Lambda}$ obtained by the different waveform models.

We discuss constraints on NS EOS models by combining information obtained from GW170817 and GW190425.
Our results using \Kyoto~model show that the chirp radius $\mathcal{R}$ is constrained between about 10 and 14 km for the chirp mass $\mathcal{M}\simeq1.2~M_\odot$.
By using an independent waveform model (\Kyoto~model) and independent analysis, 
we obtain the results consistent with the LVC's one: a low SNR event from a massive BNS like GW190425
cannot contribute very much to constraining the NS EOS as shown in Ref.~\cite{Abbott:2020uma}.
As the number of BNS merger events increases
and sensitivities of detectors are improved, the systematic differences will become significant.

%%%%%%%%%%%%%%%%%%%%%%%%%%%%%%%%%%%%%%%
\section*{Acknowledgment}
%%%%%%%%%%%%%%%%%%%%%%%%%%%%%%%%%%%%%%%
We thank John Veitch for very helpful explanation of LALInference and
C. van den Broeck for useful discus- sions. This work is supported by
Japanese Society for the Promotion of Science (JSPS) KAKENHI Grants
No. JP15K05081, No. JP16H02183, No. JP16H06341, No. JP16H06342, No. JP17H01131,
No. JP17H06133, No. JP17H06358, No. JP17H06361, No. JP17H06364, No. JP18H01213, No. JP18H04595,
No. JP18H05236, and No. JP19K14720, and by a post-K project hp180179. This work
is also supported by JSPS Core-to-Core Pro- gram A. Advanced Research
Networks and by the joint research program of the Institute for Cosmic
Ray Re- search, University of Tokyo, Computing Infrastructure Project of
KISTI-GSDC in Korea, and Computing Infrastructure ORION in Osaka
City University. T. Narikawa is supported in part by a Grant-in-Aid for
JSPS Research Fellows, and he also thanks hospitality of van den Broeck's group
during his stay at Nikhef. K. Kawaguchi was supported in part by JSPS
overseas research fellowships. We are also grateful to the LIGO-Virgo
collaborations for the public release of GW data of
GW170817. This research has made use of data, software, and web tools
obtained from the Gravitational Wave Open Science Center~\cite{GWOSC}, a service of LIGO Laboratory, the LIGO
Scientific Collaboration and the Virgo Collaboration. LIGO is funded by
the U. S. National Science Foundation. Virgo is funded by the French
Centre National de la Recherche Scientifique (CNRS), the Italian
Istituto Nazionale di Fisica Nucleare (INFN), and the Dutch Nikhef, with
contributions by Polish and Hungarian institutes.

\appendix

\begin{widetext}

%----------------------------------------------------------------%
\begin{figure}[htbp]
  \begin{center}
\begin{tabular}{cc}
 \begin{minipage}[b]{0.45\linewidth}
 \begin{center}
    \includegraphics[keepaspectratio=true,height=80mm]{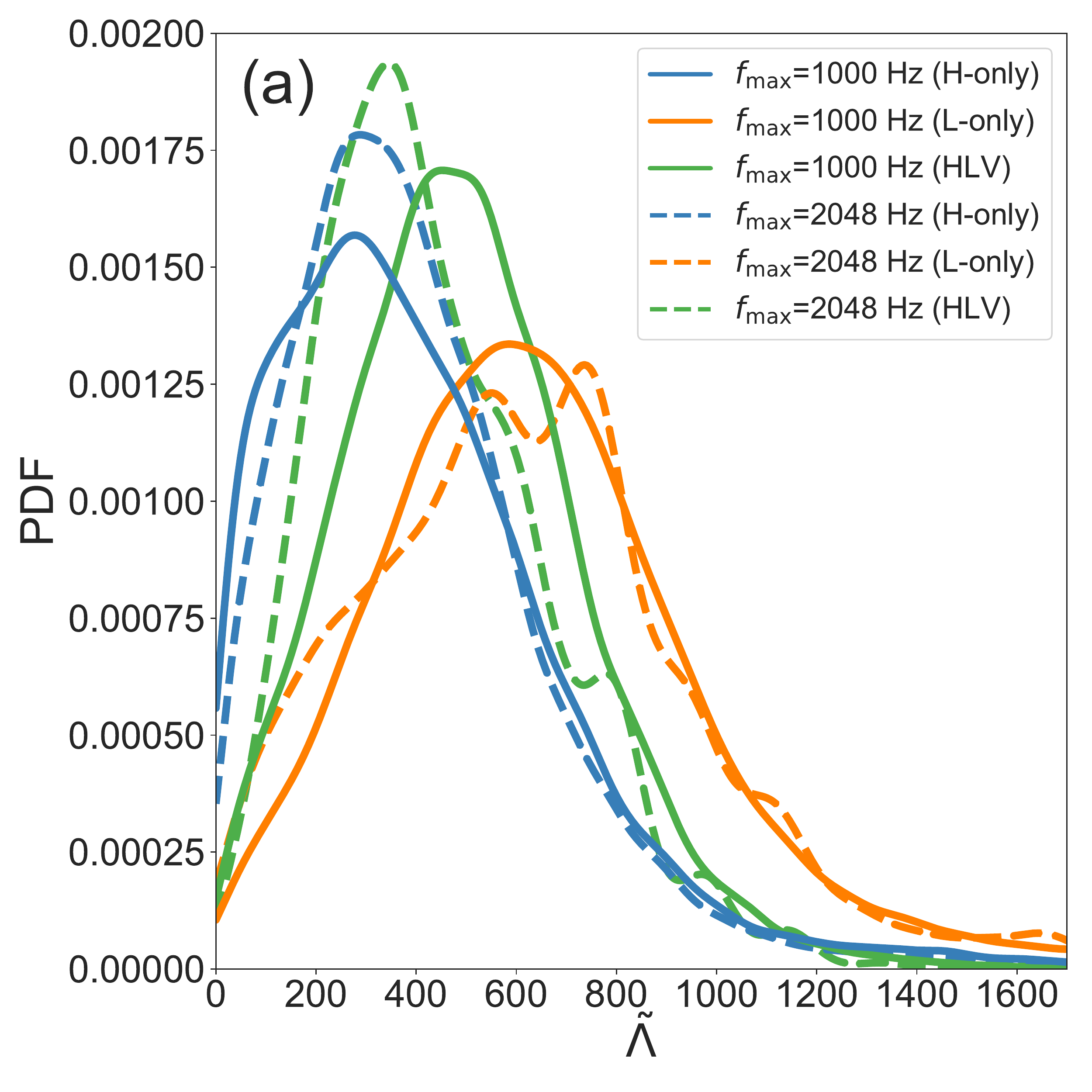}\\
 \end{center}
 \end{minipage}
 \begin{minipage}[b]{0.45\linewidth}
  \begin{center}
    \includegraphics[keepaspectratio=true,height=80mm]{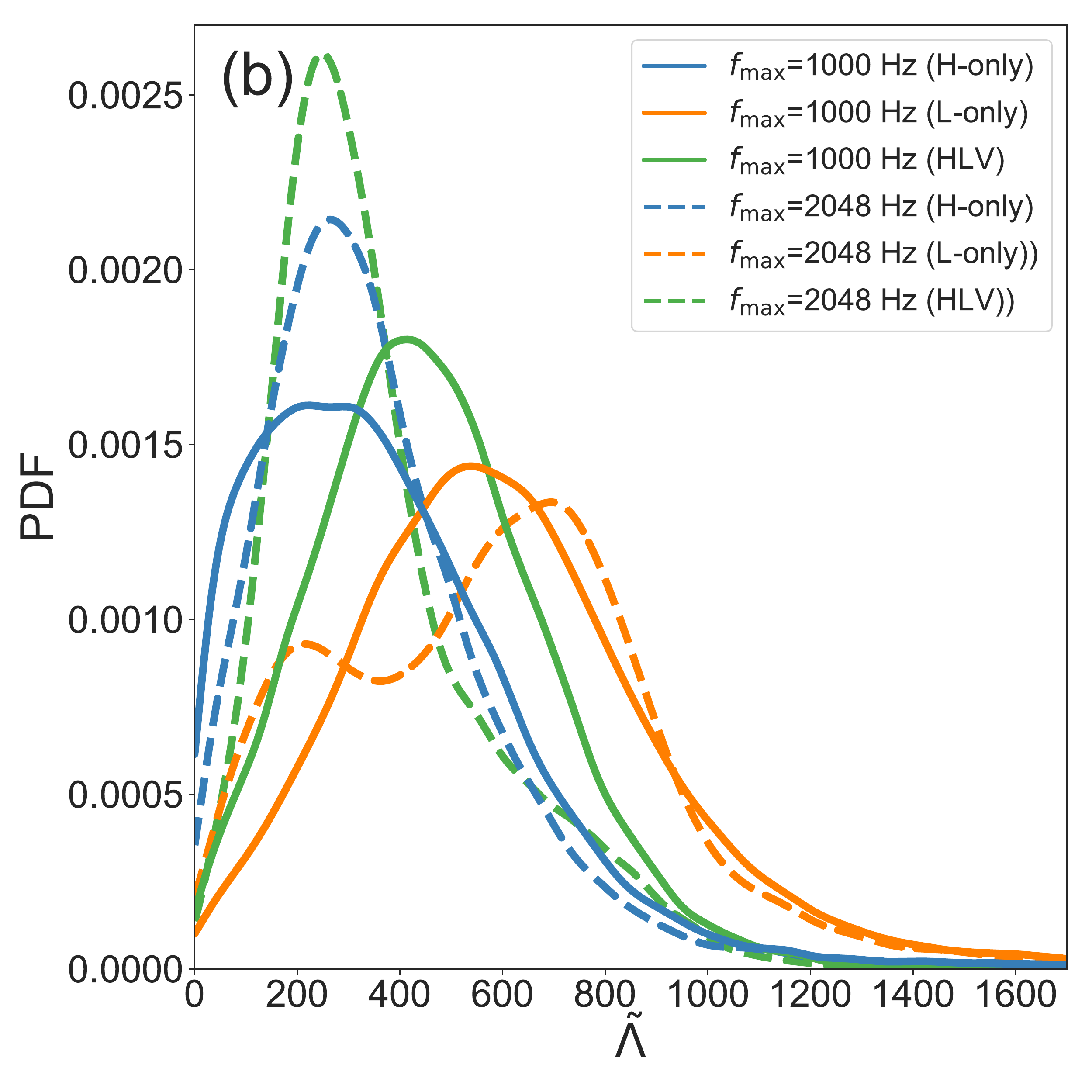}\\
 \end{center}
 \end{minipage}
\end{tabular}
  \caption{
Marginalized posterior PDFs of binary tidal deformability, $\tilde{\Lambda}$, for GW170817 derived by data of different detector combinations 
with both $f_\mathrm{max} = 1000~\mathrm{Hz}$ (solid) and  2048 Hz (dashed)
for (a) the \Kyoto~and (b) the \NRTidalvt~models.
The distribution derived by the Hanford-only data (blue), that by the Livingston-only data (orange), and that by combined data of Advanced LIGO twin detectors and Advanced Virgo (green, denoted by HLV) are presented.
For $f_\mathrm{max}=2048~\mathrm{Hz}$, 
a multimodal (bump) structure at high-$\tilde{\Lambda}$ for the \Kyoto~(\NRTidalvt) model appear
due to Livingston data.
}%
\label{fig:lamt_Kyoto_NRTidalv2}
\end{center}
\end{figure}
%----------------------------------------------------------------%

\begin{table}
 \caption{90\% credible interval of binary tidal deformability, for GW170817
 $\tilde{\Lambda}$, with the \Kyoto~(left side) and the \NRTidalvt~(right side) models, for different detector data and the maximum frequency,
 $f_\mathrm{max}$. The upper group shows the symmetric intervals and the
 lower shows the HPD intervals, where the median
 is shown as a representative value for both groups.}
 \begin{tabular}{ccccc} \hline
  & \multicolumn{2}{c}{\Kyoto} & \multicolumn{2}{c}{\NRTidalvt} \\
    $f_\mathrm{max}$ & Hanford-only & Livingston-only & Hanford-only & Livingston-only \\
  \hline \hline
  & \multicolumn{4}{c}{Symmetric} \vspace*{2pt} \\
  \SI{1000}{\hertz} & $357^{+568}_{-311}$ & $618^{+637}_{-447}$ &
              $333^{+514}_{-291}$ & $582^{+586}_{-413}$ \vspace{2pt} \\
  \SI{2048}{\hertz} & $362^{+514}_{-295}$ & $607^{+658}_{-482}$ &
              $320^{+481}_{-253}$ & $589^{+549}_{-487}$ \vspace{2pt} \\
  \hline
 &  \multicolumn{4}{c}{HPD} \vspace*{2pt} \\
  \SI{1000}{\hertz} & $357^{+414}_{-357}$ & $618^{+502}_{-523}$ &
              $333^{+378}_{-333}$ & $582^{+477}_{-484}$ \vspace{2pt} \\
  \SI{2048}{\hertz} & $362^{+378}_{-352}$ & $607^{+511}_{-557}$ &
              $320^{+355}_{-305}$ & $589^{+399}_{-555}$ \vspace{2pt} \\
  \hline
 \end{tabular}
 \label{table:lamt_Kyoto_NRTidalv2}
\end{table}

\end{widetext}

\section{Separate analysis for the LIGO twin detectors for GW170817}
\label{sec:discrepancy}
There is a multimodal structure at the high-$\tilde{\Lambda}$ region in the posterior 
PDF of $\tilde{\Lambda}$ for GW170817 for the \Kyoto~model 
and a bump structure for the \NRTidal~and \NRTidalvt~models
for $f_\mathrm{max}=2048~\mathrm{Hz}$ as shown in Fig.~\ref{fig:lamt}(b).
In this appendix, we present an in-depth study to interpret these features
by separate analysis for the LIGO twin detectors for GW170817.
Figure~\ref{fig:lamt_Kyoto_NRTidalv2} shows marginalized posterior of $\tilde{\Lambda}$ derived by separate analysis for the Hanford and Livingston detectors 
with both $f_\mathrm{max} = 1000~\mathrm{Hz}$ and 2048 Hz
for (a) the \Kyoto~model and (b) the \NRTidalvt~model.
Table~\ref{table:lamt_Kyoto_NRTidalv2} shows corresponding 90\% credible intervals of $\tilde{\Lambda}$.

In the case of the \Kyoto~model, Fig.~\ref{fig:lamt_Kyoto_NRTidalv2}(a) suggests that the origin of 
the bump at the high-$\tilde{\Lambda}$ region for $f_\mathrm{max} = 2048~\mathrm{Hz}$ for the HLV combined data is as follows.
On the one hand, for the Livingston data, the unimodal distribution for $f_\mathrm{max} = 1000~\mathrm{Hz}$, whose peak is at about 600,
is separated into a bimodal distribution for $f_\mathrm{max} = 2048~\mathrm{Hz}$ that is constructed from twin peaks, a low-$\tilde{\Lambda}$ bump, and a few high-$\tilde{\Lambda}$ bumps.
On the other hand, for the Hanford data, the unimodal distribution for $f_\mathrm{max} = 1000~\mathrm{Hz}$, whose peak is at low-$\tilde{\Lambda}$ region, shrinks for $f_\mathrm{max} = 2048~\mathrm{Hz}$.
As a result, for $f_\mathrm{max} = 2048~\mathrm{Hz}$, the remaining high-$\tilde{\Lambda}$ peak for the Livingston data produces the bump for the HLV combined data.
Moreover, a few high-$\tilde{\Lambda}$ bumps in the case of HLV combined data for $f_\mathrm{max}=2048~\mathrm{Hz}$ are inherited from the bumps of the Livingston-only data, 
which are associated with the high-frequency data.
The location of the low-$\tilde{\Lambda}$ bump derived by the Livingston-only data is close to the peak of $\tilde{\Lambda}$ of about 250 derived by the Hanford-only data.

In the case of the \NRTidalvt~model, 
as shown in Fig.~\ref{fig:lamt_Kyoto_NRTidalv2} (b), 
a bump at the high-$\tilde{\Lambda}$ region in the case of HLV combined data 
for $f_\mathrm{max}=2048~\mathrm{Hz}$
are inherited from the peak of the Livingston-only data, $\tilde{\Lambda}\sim750$.

\begin{figure}[htbp]
  \begin{center}
    \includegraphics[keepaspectratio=true,height=80mm]{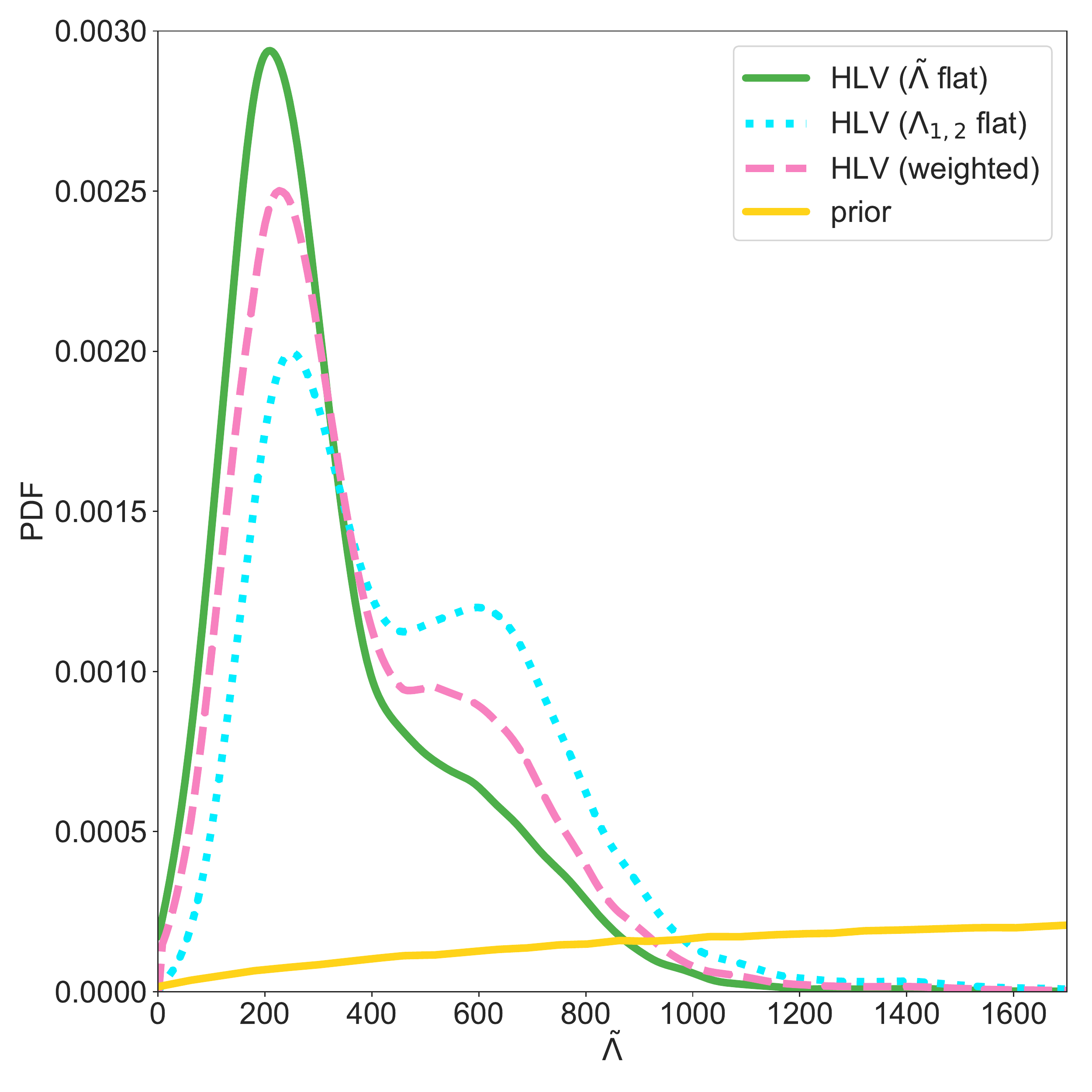}\\
  \caption{
  Dependence of the marginalized posterior PDFs of $\tilde{\Lambda}$ on different priors in $\tilde{\Lambda}$ for GW170817 for the \NRTidal~model with $f_\mathrm{max} = 2048~\mathrm{Hz}$.
 In addition to PDF of $\tilde{\Lambda}$ for uniform priors in $\Lambda_1$ and $\Lambda_2$  (dotted,  cyan),
 we show the PDF for ``Weighted''-prior (dashed, magenta), which is weighted by dividing the original prior (also shown by solid yellow curve)
 and the PDF for a uniform prior in $\tilde{\Lambda}$ (solid,  green).
}%
\label{fig:NRTidal_prior_comparison}
\end{center}
\end{figure}

While a bimodal distribution appears
in the posterior PDF of $\tilde{\Lambda}$ with the \texttt{SEOBNRv4\_ROM\_NRTidal} model
in the case of LVC analysis
as shown in Fig.~11 in \cite{Abbott:2018wiz},
a small high-$\tilde{\Lambda}$ bump at $\tilde{\Lambda}\sim600$ appears
in that with the \NRTidal~model presented for $f_\mathrm{max}=2048~\mathrm{Hz}$ in Fig.~\ref{fig:lamt} (b).
Here, the \texttt{SEOBNRv4\_ROM\_NRTidal} model is constructed from the \texttt{SEOBNRv4} model~\cite{Bohe:2016gbl,Purrer:2014fza} as the BBH baseline and
the \texttt{NRTidal} model as the tidal part.
Supplementary analysis with the \NRTidal~model as shown in Fig.~\ref{fig:NRTidal_prior_comparison}
demonstrates that the different priors in $\tilde{\Lambda}$ (one uniform and one non-uniform) make such different distribution between our analysis and the LVC analysis.
The LVC used ``Weighted'' prior.
In this prior, they assume uniform priors in $\Lambda_1$ and $\Lambda_2$, 
and weight the posterior of $\tilde{\Lambda}$ by dividing by its prior
determined by those of other parameters \cite{Abbott:2018wiz}. 
``Weighted'' prior approximately corresponds to imposing a uniform prior on $\tilde{\Lambda}$.
Figure~\ref{fig:NRTidal_prior_comparison} shows the dependence of the results on different priors in $\tilde{\Lambda}$, ``$\Lambda_{1,2}$ flat'', ``weighted,'' and ``$\tilde{\Lambda}$ flat'' for the \NRTidal~model with $f_\mathrm{max} = 2048~\mathrm{Hz}$.
This figure demonstrates that 
the distribution for ``$\Lambda_{1,2}$ flat'' and ``weighted'' prior tends to be a bimodal rather than a high-$\tilde{\Lambda}$ bump.

In Ref.~\cite{Narikawa:2018yzt}, it is found that there is 
a discrepancy in the estimates of binary tidal deformability of GW170817 between the Hanford and Livingston detectors of Advanced LIGO by using the restricted \TaylorFt~waveform model.
Figure~\ref{fig:lamt_Kyoto_NRTidalv2} shows that the discrepancy is enhanced with sophisticated waveform models (the \Kyoto~and \NRTidalvt~models).
While the two distributions in the cases of the Hanford-only and Livingston-only data seem to be consistent with each other and also consistent with what we
would expect from noise realization (see e.g., Ref.~\cite{Wade:2014vqa}), 
the results that the width of the 90\% credible interval for the Livingston-only data 
does not shrink as $f_\mathrm{max}$ increases indicate that 
the Livingston's high-frequency data are not very useful to determine the tidal deformability for GW170817.

%----------------------------------------------------------------%
%----------------------------------------------------------------%

%%%%%%%%%%%%%%%%%%%%%%%%%%%%%%%%%%%%%%%%
%---------   References   ---------%

%%%%%%%%%%%%%%%%%%%%%%%%%%%%%%%%%%%%%%%%

\end{document}